\documentclass[aps,pre, reprint]{revtex4-1}
\usepackage{times}		
\usepackage{booktabs}
\usepackage{placeins}

\usepackage{amssymb,amstext,amsmath, epsf}
\usepackage{graphicx,graphics,epsfig,wrapfig}
\usepackage{listings}
\usepackage{float}
\usepackage{esint}

\usepackage{color}
\usepackage[english]{babel}
\usepackage[T1]{fontenc}
\usepackage[utf8]{inputenc}

\def\apj{{ApJ}}
\def\apjl{{ApJL}}

\def\mnras{{MNRAS}}
\def\nat{{Nature Physics}}
\def\prl{{PRL}}

\def\pre{{PRE}}

\def\apss{{Astrophysics and Space Science}}

\begin{document}

\title{Transport of and Radiation Production by Trans-relativistic/Non-relativistic Particles Moving Through Sub-Larmor-Scale Electromagnetic Turbulence}
\author{Brett D. Keenan}
\email{bdkeenan@ku.edu}
\author{Alexander L. Ford}
\email{alexlford@ku.edu}
\affiliation{Department of Physics and Astronomy, University of Kansas, Lawrence, KS 66045}
\author{Mikhail V. Medvedev}
\altaffiliation{Also at the ITP, NRC ``Kurchatov Institute", Moscow 123182, Russia}
\affiliation{Department of Physics and Astronomy, University of Kansas, Lawrence, KS 66045}


\begin{abstract}
Plasmas with electromagnetic fields turbulent at sub-Larmor-scales are a feature of a wide variety of high-energy-density environments, and are essential to the description of many astrophysical/laboratory plasma phenomena. Radiation from particles, whether they be relativistic or non-relativistic, moving through small-scale magnetic turbulence has spectral characteristics distinct from both synchrotron and cyclotron radiation. The radiation, carrying information on the statistical properties of the magnetic turbulence, is also intimately related to the particle diffusive transport. We have investigated, both theoretically and numerically, the transport of non-relativistic and trans-relativistic particles in plasmas with high-amplitude isotropic sub-Larmor-scale magnetic turbulence, and its relation to the spectra of radiation simultaneously produced by these particles. Consequently, the diffusive and radiative properties of plasmas turbulent on sub-Larmor scales may serve as a powerful tool to diagnosis laboratory and astrophysical plasmas.
\end{abstract}

\maketitle

\section{Introduction}
\label{s:intro}

High-amplitude sub-Larmor-scale electromagnetic turbulence is a phenomenon largely associated with high-energy density environments. Such turbulence is a common feature of astrophysical and space plasmas, e.g., at high-Mach-number collisionless shocks in weakly magnetized plasmas \citep{medvedev09, frederiksen04, nishikawa03}, upstream regions of quasi-parallel shocks \citep{sironi06, plotnikov12}, sites of magnetic reconnection \citep{swisdak08, liu09} and others. Additionally, these sub-Larmor-scale, or ``small-scale'', fields play a critical role in laboratory plasmas; especially in high-intensity laser plasmas -- as observed in facilities such as the National Ignition Facility (NIF), OmegaEP, Hercules, Trident, and others \citep{ren04, huntington12, mondal12, tatarakis03}. Experimental and numerical studies of non-relativistic collisionless shocks also show that they are mediated by small-scale electromagnetic turbulence \citep{fiuza12, medvedev06}. Thus, studies of plasmas and turbulence in these environments are important for the fusion energy sciences and the inertial confinement concept \citep{ren04, tatarakis03}.

Small-scale electromagnetic turbulence can be of various origin and thus have rather different properties, from being purely magnetic (Weibel) turbulence \citep{weibel59, fried59, medvedev09c}, to various types of electromagnetic turbulence (for example, whistler wave turbulence or turbulence produced by filamentation/mixed mode instability \citep{lemoine09, bret05}), to purely electrostatic Langmuir turbulence \citep{treumann97, bret05b}.

Despite substantial differences, these small-scale fields share one thing in common: they vary on scales much smaller than the characteristic curvature scale of the particles traversing the field, i.e.\ the plasma inertial length (skin depth) which are on the order of the particle Larmor radius. The particle trajectory through these turbulent fields will, consequently, never form a well-defined Larmor circle. 

If the electromagnetic fields are random, which is usually the case of turbulence because of the random phases of fluctuations, the paths of the particles diffusively diverge due to pitch-angle diffusion. Radiation simultaneously produced by these particles is neither cyclotron nor synchrotron (for non-relativistic or relativistic particles, respectively) but, instead, carries information about the spectrum of turbulent fluctuations. Here we stress that we strictly consider the case of turbulence in vanishing mean field plasma $\langle {\bf B} \rangle = 0$.

\indent
In our previous work, see Ref. \citep{keenan13}, we found the relation between the transport of relativistic particles in isotropic three-dimensional small-scale magnetic turbulence and the radiation spectra simultaneously produced by these particles. In particular, we found that the radiation spectrum agrees with the small-angle jitter radiation prediction, in the small deflection angle regime \citep{medvedev00,medvedev06,medvedev11,RK10,TT11}. Furthermore, we demonstrated that the pitch-angle diffusion coefficient is directly related to, and can readily be deduced from, the spectra of the emitted radiation. This inter-relation between radiative and transport properties provides a unique way to remotely diagnose high-energy-density plasmas, both in laboratory experiments and in astrophysical systems. 
\newline
\indent
We extend our previous work to now consider non-relativistic ($v \lesssim 0.1c$) and trans-relativistic (i.e.\ mildly relativistic: $0.1c \lesssim v \lesssim 0.5c$) particles moving through three-dimensional sub-Larmor-scale magnetic turbulence. We demonstrate, once more via numerical and theoretical analysis, that an analogous inter-relation holds in these regimes as well, which naturally generalizes the relativistic small-angle jitter radiation regime and the pitch-angle diffusion coefficient. 
\newline
\indent
This trans-relativistic regime is applicable to laboratory plasmas, particularly high-intensity laser plasmas -- where bulk plasma motion is below $v \lesssim 0.5c$. Multi-dimensional relativistic Particle-In-Cell (PIC) simulations and laboratory experiments have revealed that non-relativistic collisionless shocks, mediated by Weibel-like instabilities, can occur in an overcritical plasma via interaction with an ultraintense laser pulse \citep{fiuza12, mondal12}. In the laboratory setting, laser-produced supersonic counter-streaming plasmas have been observed to give rise to self-organized electromagnetic fields \citep{kugland12}. Recently, the formation of filamentary structures indicative of Weibel-like magnetic fields, fully consistent with the shock model offered by 3D PIC simulations and theoretical instability analysis, has been directly observed in a scaled laboratory experiment \citep{huntington15}. Consequently, given the role of trans-relativistic particle motion in these environments, the study of the small-scale electromagnetic turbulence may be aided by the diagnostic tool offered via this inter-relation between the transport and radiative properties.
\newline
\indent
The rest of the paper is organized as follows. Section \ref{s:analytic} presents the analytic theory. Sections \ref{s:model} and \ref{s:results} describe the numerical techniques employed and the obtained simulation results. Section \ref{s:concl} is the conclusions. All equations appear in cgs units.

\section{Analytic theory}
\label{s:analytic}

\subsection{Pitch-angle diffusion}
\label{s:diffusion}

Consider a trans-relativistic electron moving (with velocity, ${\bf v}$) through a non-uniform, random, mean-free (i.e.\ $\langle {\bf B} \rangle = 0$), small-scale magnetic field (and assume that this magnetic ``micro-turbulence'' is statistically homogeneous and isotropic). Because the Lorentz force on the electron is random, it's velocity and acceleration vectors vary stochastically, leading to a random (diffusive) trajectory. We define the field turbulence to be ``small-scale'' when the electron's Larmor radius, $r_L = \gamma\beta m_e c^2/e \langle B_\perp^2 \rangle^{1/2}$ (where $\beta=v/c$ is the dimensionless particle velocity, $m_e$ is the electron mass, $c$ is the speed of light, $e$ is the electric charge, $\langle B_\perp^2 \rangle^{1/2}$ is the rms component of the magnetic field perpendicular to the electron's velocity vector, and $\gamma$ is the electron's Lorentz factor) is greater than, or comparable to, the characteristic correlation scale of the magnetic field, $\lambda_B$, i.e., $r_L\gtrsim \lambda_B$.

For small deflections, the deflection angle of the velocity (with respect to the particle's initial direction of motion) is approximately the ratio of the change in the electron's transverse momentum to its initial transverse momentum. The former is $ \sim F_L\tau_\lambda$, where ${\bf F}_L=(e/c)\,{\bf v\times B}$ is the transverse Lorentz force, and $\tau_\lambda$ is the transit time, which is the time required to traverse the scale of the field's inhomogeneity, i.e., the field correlation length, $\lambda_B$. This is, $\tau_\lambda  \sim \lambda_B/v_\perp$ -- where $v_\perp$ is the the component of the velocity perpendicular to the magnetic field. The change in the transverse momentum is thus, ${\Delta}p_\perp \sim F_L \tau_\lambda \sim e(B/c)\lambda_B$. Given that the particle's total transverse momentum is $p_\perp \sim \gamma m_e v_\perp$, the deflection angle over the field correlation length will be $\alpha_\lambda \approx {\Delta}p_\perp/p_\perp \sim e(B/c)\lambda_B/\gamma m_e v_{\perp}$. The subsequent deflection will be in a random direction, because the field is uncorrelated over scales greater than $\lambda_B$, hence the particle motion is diffusive. As for any diffusive process, the ensemble-averaged squared deviation grows linearly with time. Hence, for the pitch-angle deviation, we have
\begin{equation}
\langle \alpha^2 \rangle = D_{\alpha\alpha}t.
\label{diff_def}
\end{equation} 
The pitch-angle diffusion coefficient is, by definition, the ratio of the square of the deflection angle in a coherent patch to the transit time over this patch, that is
\begin{equation}
D_{\alpha\alpha} \sim \frac{\alpha_\lambda^2}{\tau_\lambda}\sim \left(\frac{e^2}{m_e^2 c^3}\right)\frac{1}{\langle \beta_{\perp}^2 \rangle^{1/2}}\frac{\lambda_B}{\gamma^2}{\langle B^2 \rangle},
\label{Daa}
\end{equation}
where a volume-averaged square magnetic field, $\langle B^2 \rangle$ and perpendicular rms velocity, $\langle \beta_{\perp}^2 \rangle^{1/2}$ have been substituted for $B^2$ and $\beta_{\perp} \equiv v_{\perp}/c$. Note that the diffusion coefficient depends on both statistical properties of the magnetic field, namely its strength and the typical correlation scale. 

Although the assumption that $\alpha_\lambda \ll 1$ is valid in the ultra-relativistic limit: $\beta \rightarrow 1$ (see Ref. \citep{keenan13}), it is not evident that it holds for trans-relativistic and non-relativistic velocities. As we will demonstrate via numerical simulation, pitch-angle diffusion will occur in accordance with Eq. (\ref{Daa}), so long as the magnetic turbulence is sub-Larmor-scale, i.e.\ $r_L \gtrsim \lambda_B$. 

The average square magnetic field, $\langle B^2 \rangle$ is related to $\langle B_\perp^2 \rangle$ by a multiplicative factor. For isotropic magnetic turbulence, $\langle B_x^2 \rangle = \langle B_y^2 \rangle = \langle B_z^2 \rangle$. Thus, $\frac{1}{3}\langle B^2 \rangle = \langle B_x^2 \rangle$. Alternatively, ${\bf B}$ may be expressed as a linear combination of parallel and perpendicular components. Given isotropy, $\langle B_{\perp}^2 \rangle = \langle B_x^2 \rangle + \langle B_y^2 \rangle$, so 
\begin{equation}
\langle B_\perp^2 \rangle = \frac{2}{3}\langle B^2 \rangle.
\label{b_perp}
\end{equation}
Recognizing that $v_{\perp}B = vB_{\perp}$ allows the expression of the rms perpendicular velocity as
\begin{equation}
\langle \beta_{\perp}^2 \rangle^{1/2} = \sqrt{\frac{2}{3}}\beta,
\label{beta_para}
\end{equation}
Next, the correlation length, $\lambda_B$ lacks a formal definition. It is, nonetheless, commonplace in the literature -- e.g.\ Ref. \citep{biswas02}, to define the two-point autocorrelation tensor,
\begin{equation}
R^{ij}({\bf r}, t) \equiv \langle {B}^i({\bf x}, \tau){B}^j({\bf x} + {\bf r}, \tau + t) \rangle_{{\bf x}, \tau},
\label{corr_tensor}
\end{equation}
with the formally path and time dependent correlation length tensor defined as
\begin{equation}
\lambda^{ij}_B(\hat{\bf r}, t) \equiv \int_{0}^\infty \! \frac{R^{ij}({\bf r}, t)}{R^{ij}(0, 0)}  \, \mathrm{d}r.
\label{corr_l}
\end{equation}
Note that we make no distinction between co-variant and contra-variant components; the usage of upper and lower indices is only for convenience.

Since the component of the magnetic field perpendicular to the particle trajectory alters the motion, we choose an integration path along ${\bf v_\perp}$ and only consider a transverse magnetic field component. In accord with standard practice (see, for example, Ref. \citep{batchelor82}), we choose ${\bf r} = z\hat{\bf z}$ and $i=j=x$. Thus, we define the magnetic field correlation length as
\begin{equation}
\lambda_B \equiv \lambda^{xx}_B(\hat{\bf z}, t)  = \int_{0}^\infty \! \frac{R^{xx}(z\hat{\bf z}, t)}{R^{xx}(0, 0)}  \, \mathrm{d}z.
\label{corr_l_def}
\end{equation}
The correlation length has a convenient representation in Fourier ``$k$-space'' and ``$\Omega$-space". Let ${\bf B}_{{\bf k}, \Omega}$ be the spatial and temporal Fourier transform of the magnetic field, i.e.\
\begin{equation}
{\bf B}_{{\bf k},\Omega} = \int \! {\bf B}({\bf x}, t)e^{-i({\bf k}\cdot{\bf x} - \Omega{t})} \, \mathrm{d} {\bf x} \mathrm{d}t,
\label{B_fourier_def}  
\end{equation} 
where ${\bf k}$ and $\Omega$ are the corresponding wave vector and frequency, respectively. We may define a complementary spectral correlation tensor $\Phi_{ij}({\bf k}, \Omega)$, such that
\begin{equation}
R_{ij}({\bf r}, t) = (2\pi)^{-4}\int \Phi_{ij}({\bf k}, \Omega) e^{i{\bf k}\cdot{\bf r} -i\Omega{t}} \, \mathrm{d}{\bf k}\, \mathrm{d}{\Omega},
\label{corr_tensor_def_Phi}
\end{equation}
Isotropy, homogeneity, time-independence, and ${\bf \nabla}\cdot{\bf B} = 0$ require that the spectral correlation tensor take the simple form \citep{biswas02}
\begin{equation}
\Phi_{ij}({\bf k}, \Omega) = \frac{1}{2V}\left|{\bf B}_k\right|^2\left(\delta_{ij} - \hat{k}_i\hat{k}_j\right)2\pi\delta(\Omega),
\label{spec_tensor}
\end{equation}
where $V$ is the volume of the space considered, $\hat{\bf k}$ is the unit vector in the direction of the wave vector, and $\delta_{ij}$ is the Kronecker delta. The normalization has been chosen such that $\sum{R}_{ii}(0, 0) = \langle B^2 \rangle_{{\bf x}, \tau} = \langle B^2 \rangle$. Given Eq. (\ref{corr_tensor_def_Phi}) and Eq. (\ref{spec_tensor}), the correlation length may be reformulated as
\begin{equation}
\lambda_B = \int_{0}^\infty \!  \frac{\int \! |{\bf B}_k|^2k^{-2}(k^2-k_x^2)e^{ik_z{z}}\, \mathrm{d}{\bf k}}{\int \! |{\bf B}_k|^2k^{-2}(k^2-k_x^2)\, \mathrm{d}{\bf k}} \, \mathrm{d}{z}.
\label{corr_l}
\end{equation}
By assuming isotropic turbulence, the magnetic field has azimuthal and polar symmetry in $k$-space, hence ${\bf B}_{\bf k}$ is only a function of $|{\bf k}| \equiv k$. After the integration over $z$ and all solid-angles in Fourier space, Eq. (\ref{corr_l}) becomes
\begin{equation}
\lambda_B = \frac{3\pi}{8}\frac{\int_{0}^\infty \! k{|{\bf B}_k|^2}\, \mathrm{d}k}{\int_{0}^\infty \! k^2{|{\bf B}_k|^2}\, \mathrm{d}k}.
\label{corr_l_div}
\end{equation}
It may be noted that $\lambda_B \approx k_B^{-1}$, where $k_B$ is the characteristic (dominant) wave number of turbulence. 

Thus, with Eqs. (\ref{Daa}), (\ref{beta_para}), and (\ref{corr_l_div}), the pitch-angle diffusion coefficient is
\begin{equation}
D_{\alpha\alpha} \equiv \frac{3\pi}{8}\sqrt{\frac{3}{2}}\left(\frac{e^2}{m_e^2 c^3}\right)\frac{\int_{0}^\infty \! k{|{\bf B}_k|^2}\, \mathrm{d}k}{\int_{0}^\infty \! k^2{|{\bf B}_k|^2}\, \mathrm{d}k}\frac{\langle B^2 \rangle}{\gamma^2\beta}.
\label{Daa_def}
\end{equation}
To continue, we must specify a magnetic spectral distribution, $|{\bf B}_k|^2$. As in our previous work (Ref. \citep{keenan13}), we assume the isotropic three-dimensional magnetic turbulence has a static, i.e.\ time-independent, power law turbulent spectrum:
\begin{equation}
\left\{\begin{array}{ll}
|{\bf B}_{\bf k}|^2 = Ck^{-\mu}, & k_{min} \le k \le k_{max} 
 \\
|{\bf B}_{\bf k}|^2 = 0. & \text{otherwise}
\end{array}\right.
\label{Bk}
\end{equation} 
Here the magnetic spectral index, $\mu$ is a real number, and
\begin{equation}
C \equiv \frac{2\pi^2V\langle B^2 \rangle}{\int_{k_\text{min}}^{k_\text{max}} \! k^{-\mu+2}\, \mathrm{d}k},
\label{C_def}
\end{equation} 
is a normalization, such that 
\begin{equation}
V^{-1}\int \! {\bf B}^2({\bf x}) \mathrm{d}{\bf x} = (2\pi)^{-3}\int \! |{\bf B}_{\bf k}|^2 \, \mathrm{d} {\bf k}.
\label{C_def_cont}  
\end{equation} 

It should be noted that our principal results strictly apply only to static turbulence. One should, in principle, consider time-dependent fields as well. However, if the transit time of a particle over a correlation length is shorter than the field variability time-scale, then the static field approximation is valid. Additionally, plasma instabilities generally produce random fields in a preferred direction, leading to anisotropic turbulence. Nonetheless, isotropy may arise in an advance stage of development. Magnetic turbulence of this kind is a natural outcome of the non-linear Weibel-filamentation instability, which occurs at relativistic collisionless shocks and in laser-produced plasmas \citep{medvedev00, medvedev06, medvedev11}.

\subsection{The ultra-relativistic jitter theory}
\label{s:rel_jit}

Now we consider the radiative properties of these sub-Larmor-scale plasmas. First, the ultra-relativistic radiation regime in sub-Larmor-scale magnetic turbulence is well understood. This regime is characterized by a single parameter, the ratio of the deflection angle, $\alpha_\lambda$ to the relativistic beaming angle, $\Delta\theta \sim 1/\gamma$. The ratio \citep{medvedev00, medvedev11, keenan13}
\begin{equation}
\frac{\alpha_\lambda}{\Delta\theta} \sim \frac{eB_{\perp}\lambda_B}{m_e c^2} \sim 2\pi \frac{e \langle B^2 \rangle^{1/2}}{m_e c^2 k_B} \equiv \delta_j
\label{delta}
\end{equation} 
is known as the $\emph{jitter parameter}$. From this, we recover four distinct radiation regimes. Firstly, if $\delta_j\to\infty$, the regime is the classical synchrotron radiation regime; the particle orbits are circular in the plane orthogonal to a perfectly homogeneous magnetic field. With $\delta_j>\gamma$, the regime is very similar to synchrotron, but the particle's guiding center is slowly drifting, due to slight inhomogeneity in the magnetic field. The produced spectrum is well represented by the synchrotron spectrum, and it evolves slowly in time due to the particle diffusion through regions of differing field strength. This regime may be referred to as the diffusive synchrotron regime. 

Thirdly, when $1<\delta_j<\gamma$, the particle does not complete its Larmor orbit because the $B$-field varies on a shorter scale. In this case, an onlooking observer would see radiation from only short intervals of the particle's trajectory (i.e., whenever the trajectory is near the line-of-sight), as in synchrotron, but these intervals are randomly distributed. This is the case of the large-angle jitter regime. The radiation is similar to synchrotron radiation near the spectral peak and above, but differs significantly from it at lower frequencies, see Ref. \citep{medvedev11} for details. 

Finally, If $\delta_j \ll 1$, a distant observer on the line-of-sight will see the radiation along, virtually, the entire trajectory of the particle (which will be approximately straight with small, random, transverse deviations). This is known as small-angle jitter radiation \citep{medvedev00, medvedev06, medvedev11}. The resulting radiation markedly differs from synchrotron radiation, although the total radiated power of radiation, $P_\text{tot}\equiv dW/dt$, produced by a particle in all these regimes, e.g., jitter and synchrotron, is identical: 
\begin{equation}
P_\text{tot} = \frac{2}{3} r_e^2c \gamma^2 \langle B_\perp^2 \rangle,
\label{P_tot_rel}
\end{equation}
where $r_e = e^2/m_e c^2$ is the classical electron radius.

For ultra-relativistic electrons, the radiation spectra are wholly determined by $\delta_j$ and the magnetic spectral distribution. It has been shown \citep{medvedev06, medvedev11,RK10,TT11} that monoenergetic relativistic electrons in the sub-Larmor-scale magnetic turbulence given by Eq. (\ref{Bk}) produce a flat angle-averaged spectrum below the spectral break and a power-law spectrum above the break, that is
\begin{equation}
P(\omega) \propto 
\left\{\begin{array}{ll}
\omega^0, &\text{if}~ \omega<\omega_j, \\
\omega^{-\mu + 2}, &\text{if}~ \omega_j<\omega<\omega_b, \\
0, &\text{if}~ \omega_b<\omega,
\end{array}\right.
\label{Pomega}
\end{equation}
where the spectral break is
\begin{equation}
\omega_j =\gamma^2 k_\textrm{min} c, 
\label{omegaj-kmin}
\end{equation}
which is called the jitter frequency. Similarly, the high-frequency break is 
\begin{equation}
\omega_b =\gamma^2 k_\textrm{max} c.
\label{omegab}
\end{equation}

\subsection{Non-relativistic jitter radiation}
\label{s:nonrel_jit}

In contrast, radiation from non-relativistic particles is not beamed along a narrow cone of opening angle, $\Delta\theta$. The jitter parameter is, consequently, without meaning in the non-relativistic radiation regime. Instead, the ``dimensionless scale'' (or ``gyro-number''), i.e.\ $r_L\lambda_B^{-1}$, is the only meaningful parameter:
\begin{equation}
r_L\lambda_B^{-1} \sim k_B{r_L} = k_B\frac{{\gamma}m_e{v}c}{e\langle B^2 \rangle^{1/2}} \equiv \rho,
\label{scale_para}
\end{equation} 
Given the magnetic spectral distribution exhibited by Eq. (\ref{Bk}), $k_B \sim k_\text{min}$, so
\begin{equation}
\rho = k_\text{min}r_L.
\label{rho}
\end{equation} 
\indent
As we shall see below, the radiation spectrum in this regime markedly differs from the single-harmonic cyclotron spectrum. We call this radiation ``pseudo-cyclotron'' radiation or ``non-relativistic jitter'' radiation.

\indent
Regardless of the regime, the radiation spectrum (which is the radiative spectral energy, $dW$ per unit frequency, $d\omega$, and per unit solid-angle, $d\eta$) seen by a distant observer is obtained from the equation \citep{landau75,jackson99}
\begin{equation}
\frac{d^2W}{d\omega\, d\eta} = 
\frac{e^2}{4\pi^2 c}  \left|\int_{-\infty}^\infty \! {\bf A}_{\bf k}(t)e^{i\omega{t}}\, \mathrm{d} t
\right|^2,
\label{LW}  
\end{equation} 
where
\begin{equation}
{\bf A}_{\bf k}(t) \equiv \frac{\hat{\bf n}\times[(\hat{\bf n} - {\boldsymbol\beta}) \times \dot{\boldsymbol\beta} ]}{(1 - \hat{\bf n}\cdot{\boldsymbol\beta})^2}e^{-i{\bf k}\cdot {\bf r}(t)}.
\label{A_k}  
\end{equation} 
In this equation, ${\bf r}(t)$ is the particle's position at the retarded time $t$, ${\bf k} \equiv \hat{\bf n}\omega/c$ is the wave vector which points along $\hat{\bf n}$ from ${\bf r}(t)$ to the observer and $\dot{\boldsymbol\beta} \equiv \text{d}{\boldsymbol\beta}/\text{d}t$. Since the observer is distant, $\hat{\bf n}$ is approximated as fixed in time to the origin of the coordinate system. This fully relativistic equation is obtained from the Li\'{e}nard-Wiechart potentials.  If $v \ll c$, Eq. (\ref{LW}) simplifies to
\begin{equation}
\frac{d^2W}{d\omega\, d\eta} = 
\frac{e^2}{4\pi^2 c}  \left|\int_{-\infty}^\infty \!    \ \hat{\bf n}\times(\hat{\bf n}  \times \dot{\boldsymbol\beta}) e^{i\omega{t}}\, \mathrm{d} t
\right|^2,
\label{LW_nonrel}  
\end{equation} 

Next, integrating Eq. (\ref{LW_nonrel}) over all solid-angles gives the radiated energy per frequency 
\begin{equation}
\frac{dW}{d\omega} = \frac{2{e^2}}{3\pi{c^3}}\left|{\bf w}_{\omega}\right|^2,
\label{dWdw}  
\end{equation} 
where ${\bf w}_{\omega}$ is the Fourier component of the electron's acceleration with frequency, $\omega$. Eq. (\ref{dWdw}), valid for $v \ll c$, is known as the dipole approximation \citep{landau75}. This expression may also be obtained from the Larmor formula, i.e.\
\begin{equation}
P_\text{tot} = \frac{2}{3}\frac{e^2}{c^3}|{\bf w}|^2,
\label{P_tot_nonrel}
\end{equation}
using the identity \citep{landau75}:
\begin{equation}
\frac{1}{2}\int_{-\infty}^\infty \!   |{\bf w}(t)|^2\, \mathrm{d} t = (2\pi)^{-1}\int_{0}^\infty \!   |{\bf w}_\omega|^2\, \mathrm{d} \omega.
\label{pow_identity}
\end{equation}

To proceed further, we use our previous assumption that the particle deflection angle over a field correlation length is small (i.e.\ $\alpha_\lambda \ll 1$). This condition implies the validity of the ``perturbative'' approach, whereby the particle trajectory is approximated as a straight line. For a particle moving in a magnetic field, $\left|{\bf w}_{\omega}\right|^2$ is given by the Lorentz force. In this limiting case of small deflections, we may write 
\begin{equation}
\left|{\bf w}_{\omega}\right|^2 = \left(\frac{e\beta}{m_e}\right)^2\left(\delta_{ij} - \hat{v}_i\hat{v}_j\right)B^{i*}_\omega{B^j_\omega},
\label{w_omega}  
\end{equation} 
where ${{\bf B}_\omega}$ is the temporal variation of the magnetic field along the trajectory of the electron, i.e.\
\begin{equation}
{{\bf B}_\omega} = (2\pi)^{-4} \int \!  e^{i\omega{t}} \, \mathrm{d}t \int \! {\bf B}_{{\bf k},\Omega}e^{i{\bf k}\cdot{\bf r}(t) - i\Omega{t}} \, \mathrm{d} {\bf k} \mathrm{d}\Omega.
\label{B_omega_def}  
\end{equation} 
Since the trajectory is approximately straight, ${\bf r}(t) \approx {\bf r_0} + {\bf v}t$, consequently
\begin{equation}
{{\bf B}_\omega} = (2\pi)^{-4}  \int \! e^{i{\bf k}\cdot{\bf r}_0} {\bf B}_{{\bf k}, \Omega}   \, \mathrm{d} {\bf k} \, \mathrm{d} {\Omega} \int \! e^{i(\omega+{\bf k}\cdot{\bf v} - \Omega)t } \, \mathrm{d}t,
\label{B_omega_trans}  
\end{equation} 
After the time integration, this becomes
\begin{equation}
{{\bf B}_\omega} = (2\pi)^{-3} \int \!  \delta(\omega + {\bf k}\cdot{\bf v} - \Omega)e^{i{\bf k}\cdot{\bf r}_0} {\bf B}_{{\bf k}, \Omega} \, \mathrm{d} {\bf k} \, \mathrm{d}{\Omega}.
\label{B_omega}  
\end{equation} 
Now, since the magnetic turbulence is assumed to be homogeneous (at least over a time scale greater than the particle transit time) the product of $B^{i*}_\omega{B^j_\omega}$ along a particular trajectory starting at ${\bf r}_0$ is representative of the magnetic field as a whole \citep{medvedev11}. Thus, we may consider only the volume-average of $B^{i*}_\omega{B^j_\omega}$. Performing the integration leads to
\begin{equation}
\left<B^{i*}_{\omega}B^{j}_{\omega}\right>_{{\bf r}_0} = (2\pi)^{-3}{V}^{-1} \int \! \delta(\omega + {\bf k}\cdot{\bf v} - \Omega){B}^{i}_{{\bf k}, \Omega}{B}^{j*}_{{\bf k}, \Omega} \, \mathrm{d} {\bf k} \, \mathrm{d} {\Omega}.
\label{cor_avg}  
\end{equation} 
The quantity, $B^{i*}_{{\bf k}, \Omega}B^{j}_{{\bf k}, \Omega}$, is proportional to the Fourier image of the two-point auto-correlation tensor -- i.e.\ Eq. (\ref{spec_tensor}). Thus, with Eqs. (\ref{dWdw}), (\ref{w_omega}), (\ref{cor_avg}), and (\ref{spec_tensor}), the angle-averaged radiation spectrum of a non-relativistic electron moving in static, statistically homogeneous and isotropic sub-Larmor-scale magnetic turbulence is 
\begin{equation}
\frac{dW}{d\omega} = 
\left(\frac{Tr_e^2\beta^2c}{12\pi^3V}\right)\int \! \delta(\omega + {\bf k}\cdot{\bf v})\left[1 + \left(\hat{\bf k}\cdot\hat{\bf v}\right)^2\right] \left|{\bf B}_k\right|^2\mathrm{d}{\bf k},
\label{nonrel_analy}  
\end{equation} 
where $T$ is the duration of the observation, and where we have used
\begin{equation}
\delta(\omega + {\bf k}\cdot{\bf v}) = \int \! \delta(\omega + {\bf k}\cdot{\bf v} - \Omega)\delta(\Omega) \, \mathrm{d} {\Omega}.
\label{delta_identity}  
\end{equation} 
We see that the radiation spectrum is fully determined by the magnetic spectral distribution, $\left|{\bf B}_k\right|^2$. It is instructive to consider one of the simplest such distributions -- the isotropic spectrum of a magnetic field at a single scale, $k_B$, i.e.\ 
\begin{equation}
|{\bf B}_{\bf k}|^2 = (2\pi)^3V\langle{B^2}\rangle\frac{\delta(k-k_B)}{4\pi{k_B^2}}.
\label{Bk_single}
\end{equation} 
Substitution of Eq. (\ref{Bk_single}) into Eq. (\ref{nonrel_analy}) produces the radiation spectrum
\begin{equation}
\frac{dW}{d\omega} = 
\left\{\begin{array}{ll}
\frac{T}{3k_B}r_e^2\beta\langle B^2 \rangle \left(1 + \frac{\omega^2}{\omega_\textrm{jn}^2}\right), &\text{if}~ \omega \leq \omega_\textrm{jn}
 \\
0, &\text{if}~ \omega > \omega_\textrm{jn},
\end{array}\right.
\label{nonrel_single} 
\end{equation} 
where $\omega_\textrm{jn} = k_\textrm{B}v$. Given the magnetic spectral distribution of Eq. (\ref{Bk}), the corresponding non-relativistic jitter spectrum, is
\begin{equation}
\frac{dW}{d\omega} \propto 
\left\{\begin{array}{ll}
A + D\omega^2, &\text{if}~ \omega \leq \omega_\textrm{jn}
 \\
F\omega^{-\mu+2} + G\omega^2 + K, &\text{if}~ \leq \omega_\textrm{bn} 
 \\
0, &\text{if}~  \omega > \omega_\textrm{bn},
\end{array}\right.
\label{analy_spec}  
\end{equation} 
where $\mu \neq 2$ and
\begin{equation}
A \equiv  \frac{v}{2-\mu}\left(k_\text{max}^{-\mu+2}-k_{min}^{-\mu+2}\right),
\label{A_def}  
\end{equation} 
\begin{equation}
 D \equiv -\frac{1}{v\mu}\left(k_\text{max}^{-\mu}-k_\text{min}^{-\mu}\right),
\label{D_def}  
\end{equation} 
\begin{equation}
 F \equiv \frac{v^\mu}{v}\left(\frac{1}{\mu-2} + \frac{1}{\mu}\right),
\label{F_def}  
\end{equation} 
\begin{equation}
 G \equiv  - \frac{v}{\mu}k_\text{max}^{-\mu},
\label{G_def}  
\end{equation} 
\begin{equation}
 K \equiv \frac{v}{2-\mu}k_\text{max}^{-\mu+2},
\label{K_def}  
\end{equation} 
with the jitter frequency given by the characteristic, and largest, spatial scale
\begin{equation}
\omega_\textrm{jn} = k_\textrm{min}v.
\label{omega_jn_nonrel}  
\end{equation} 
Finally, the break frequency is indicated by the smallest spatial scale, i.e.\ the maximum wave number
\begin{equation}
\omega_\textrm{bn} = k_\textrm{max}v.
\label{omega_bn_nonrel}  
\end{equation} 
Notice the structural similarity between the spectrum at frequencies less than $\omega_\textrm{jn}$ and the delta function spectrum in Eq. (\ref{nonrel_single}).

Next, the total radiated power may be obtained by integrating Eq. (\ref{nonrel_analy}) over all frequencies and dividing by the total observation time, yielding
\begin{equation}
P_{tot} = \frac{2}{3}r_e^2\beta^2c\langle{B_\perp^2}\rangle,
\label{nonrel_power}  
\end{equation} 
where we have used Eq. (\ref{b_perp}). Compare this to the total power radiated by a non-relativistic electron moving through a uniform magnetic field, 
\begin{equation}
P_{tot} = \frac{2}{3}r_e^2\beta^2cB_\perp^2,
\label{cyclo_spec}  
\end{equation} 
which follows directly from Eq. (\ref{P_tot_nonrel}). Evidently, the total power of non-relativistic jitter radiation is identical to the total power of cyclotron radiation -- with $B^2 \rightarrow \langle B^2 \rangle$; this is exactly analogous to the relation between synchrotron and relativistic jitter radiation.

The radiation spectrum, generalized to any velocity, may be obtained by a formal Lorentz transformation to the electron rest frame. Consider a relativistic electron moving with velocity $\beta$ in the (unprimed) laboratory frame. By employing the Lorentz invariant phase space volume, $d^3k/\omega(k)$ -- the radiation spectra between the two frames can readily be related by the equality \citep{jackson99} 
\begin{equation}
\frac{1}{\omega^2}\frac{d^2W}{d\omega{d\eta}} = \frac{1}{\omega'^2}\frac{d^2W'}{d\omega'{d\eta'}}.
\label{spec_invar}  
\end{equation} 
Thus, the angle-averaged laboratory radiation spectrum is obtained by integration over all solid-angles (in the lab frame) of the electron rest frame spectrum, i.e.\
\begin{equation}
\frac{dW}{d\omega} = \int \!  \frac{\omega^2}{\omega'^2}\frac{d^2W'}{d\omega'{d\eta'}} \, \mathrm{d} {\eta}.
\label{spec_def}  
\end{equation} 
We consider, once more, that the electron moves along a straight path, experiencing only small deviations in its trajectory. Consequently, we consider a Lorentz boost of the laboratory coordinates along the trajectory (z-axis). In the electron's rest frame, the field turbulence has both a time-dependent magnetic and electric component. However, since the electron is at rest in this frame, only the electric field contributes to the instantaneous particle acceleration. Via Lorentz transformation of the laboratory magnetic field, the co-moving electric field is simply 
\begin{equation}
{\bf E}'({\bf x}', t') = \gamma{\boldsymbol\beta}\times{\bf B}({\bf r}),
\label{rest_electric}  
\end{equation} 
where ${\bf r}(t) = {\bf r_0} + {\bf v}t$. Since the electron is instantaneously at rest in this frame, we may choose ${\bf x}' = 0$; thus, $t = \gamma{t'}$. The corresponding equation of motion, for the electron, is then
\begin{equation}
m_e{\bf w}'(t') = e{\bf E}'(0, t') = e\gamma{\boldsymbol\beta}\times{\bf B}({\bf r}).
\label{rest_eom}  
\end{equation} 
As before, the radiation spectrum in the rest frame is given by the Dipole approximation, Eq. (\ref{LW_nonrel}). Substitution of these results into Eq. (\ref{spec_def}) leads to
\begin{equation}
\frac{dW}{d\omega} = \frac{e^2}{4\pi^2\gamma^2c^3} \int \! \frac{\left|{\bf w}'_{\omega'}\right|^2sin^2\Theta'}{(1 - \beta{cos\theta})^2}\, \mathrm{d} {(cos\theta)} \, \mathrm{d} {\phi},
\label{spec_def_sub}  
\end{equation} 
where $\Theta'$ is the angle between the wave and acceleration vectors in the electron rest frame, and we have used the relativistic Doppler formula $\omega' = \gamma\omega(1 - \hat{\bf n}\cdot{\boldsymbol\beta})$. Next, given the equivalent form of Eq. (\ref{rest_eom}) to the lab frame equation of motion, Eq. (\ref{w_omega}), the acceleration term is given by the non-relativistic jitter spectrum with the substitution, $\omega' \rightarrow \omega'/\gamma = \omega(1 - \beta{cos\theta})$. 

The final task is to perform the integration. However, the angle $\Theta'$ must first be related to the laboratory $\theta$ and $\phi$ coordinates -- which are derived from the angle between the wave vector and the velocity, and the azimuthal angle with respect to the boost axis, respectively. With a transverse acceleration, these angles are related by \citep{rybicki}
\begin{equation}
sin^2{\Theta}' = 1 - \frac{sin^2\theta{cos}^2\phi}{\gamma^2(1 - \beta{cos\theta})^2},
\label{angle_transform}  
\end{equation} 
with $\phi' = \phi$. Thus, the angle-averaged (velocity-independent) jitter spectrum is given by the following integration of the non-relativistic jitter spectrum 
\begin{equation}
\frac{dW}{d\omega} = \frac{3}{8\gamma^2} \int_{-1}^1 \! \mathrm{d} {x} \left[\frac{1}{(1 - \beta{x})^2} + \frac{(x-\beta)^2}{(1-\beta{x})^4}\right]I(\omega_0),
\label{jitter_vel_free}  
\end{equation} 
where $I(\omega_0)$ is the non-relativistic jitter spectrum, e.g.\ Eq. (\ref{nonrel_analy}), evaluated at $\omega_0 \equiv \omega(1 - \beta{x})$. This result leads to the traditional, ultra-relativistic, jitter spectrum in the limit of $\beta \rightarrow 1$ (or, equivalently, $\gamma \rightarrow \infty$). In the trans-relativistic regime, the characteristic frequencies, Eqs. (\ref{omegaj-kmin}) and (\ref{omegab}), generalize to
\begin{equation}
\omega_\textrm{jn} \equiv \gamma^2k_\textrm{min}v,
\label{omega_jn}  
\end{equation}
and
\begin{equation}
\omega_\textrm{bn} \equiv \gamma^2k_\textrm{max}v,
\label{omega_bn}  
\end{equation} 
which are the (trans-relativistic) jitter and break frequencies, respectively. It is noteworthy that $\omega_{bn}$ is not a proper break frequency in the mildly relativistic regime. The spectrum quickly falls to zero following $\omega_{bn}$; however, the drop is not instantaneous (as it is in the ultra-relativistic limit). In the trans-relativistic regime, $\gamma \simeq 1$, of course. With this in mind, and for the sake of convenience, we retain the $\text{n}$ subscript for both the trans-relativistic and non-relativistic expressions.

From Eqs. (\ref{analy_spec}), (\ref{jitter_vel_free}), and (\ref{Daa_def}), we see that an inter-relation between the diffusive and radiative properties of trans-relativistic/non-relativistic plasmas with sub-Larmor-scale magnetic turbulence exists. Furthermore, this inter-relation owes its existence to the statistical properties of the magnetic turbulence (e.g.\ $\langle B^2 \rangle$ and $\lambda_B$). We note, however, that our radiation treament assumes small deflections; an assumption that allowed the use of the, so called, perturbation theory. Recent work (see Ref. \citep{kelner13}) has considered a formal treatment of the perturbation theory that exclusively requires that the deflection angle over a correlation length is small, i.e.\ $\alpha_\lambda \ll 1$. Due to continued diffusive scatterings of the electron, its path will eventually deviate strongly from its initial trajectory. The traditional perturbative approach, regardless, remains valid so long as the trajectory remains approximately straight over the radiation formation length, at least for the considered domain of frequencies (i.e.\ lower frequencies will, inevitably, require a non-perturbative treatment). In the non-relativistic limit, the formation length is $\sim k^{-1}$. This must be less than, or comparable to, the magnetic correlation length $\lambda_B$. With the characteristic frequency $\omega_{jn}$, this length is $\sim \lambda_B/\beta$. Consequently, as long as the particle velocity is not arbitrary small, the perturbative approach will be valid; if $\alpha_\lambda$ is, indeed, small. By way of numerical simulation, we will demonstrate that this condition holds as long as $\rho > 1$ (i.e.\ the turbulence is sub-Larmor in scale).

Finally, our results do not consider the dispersive effect of the surrounding plasma. An account of dispersion will modify the radiation spectrum by a multiplication of Eq. (\ref{dWdw}) by the square root of the frequency-dependent scalar permittivity, $\epsilon(\omega)$. The scalar dielectric permittivity at high frequencies is \citep{jackson99, rybicki}
\begin{equation}
\epsilon(\omega) = 1 - \frac{\omega_\text{pe}^2}{\omega^2},
\label{epsilon_def}  
\end{equation} 
where $\omega_\text{pe}$ is the plasma frequency. Eq. (\ref{epsilon_def}) holds formally for $\omega^2 \gg \omega_\text{pe}^2$ in any dielectric medium; although it holds for cold, non-magnetized, isotropic plasmas for a wide domain of frequencies -- including $\omega < \omega_\text{pe}$ \citep{rybicki}. In a magnetized plasma, additional terms including the ambient ``mean'' magnetic field appear in the permittivity tensor. As previously mentioned, the Weibel-like magnetic turbulence can occur in a non-magnetized environment, thus we ignore any ``mean'' field here. Hence, we will consider an extension of Eq. (\ref{epsilon_def}) to low frequencies ($\omega \sim \omega_\text{pe}$).

The plasma dispersion effect is only important for frequencies $\omega \ll \gamma\omega_\text{pe}$ -- below which, suppression of relativistic beaming (due to the Razin effect) occurs \citep{jackson99, rybicki}. Electron driven Weibel-like turbulence occurs on a very small-scale, with $\lambda_B \sim d_e$ (where $d_e \equiv c/\omega_\text{pe}$ is the electron skin depth) \citep{medvedev09c, TT11}. Consequently, in the ultra-relativistic regime, the jitter frequency is many orders of magnitude larger than the plasma frequency -- by a factor $\sim \gamma^2$. However, in the non-relativistic and trans-relativistic regimes, dispersion can play a considerable role. This will especially be so for $\beta \ll 1$. In this case, a considerable portion of the radiation spectrum may fall below $\omega_\text{pe}$, and thus be unobservable. For simplicity and convenience, we have ignored the plasma dispersion in our numerical simulations. However, we consider a few cases with plasma dispersion intact, both numerically and theoretically, in Appendix \ref{s:appendixc}. 

\section{Numerical model}
\label{s:model}

Using the method from our previous work (see Ref. \citep{keenan13}), here we explore the inter-relation between the diffusive and radiative properties of these plasmas, and thereby test our theoretical predictions. As before, this was done via simulations of particle dynamics in sub-Larmor-scale magnetic turbulence. In our simulations, only first-principles were used. Non-relativistic and trans-relativistic electrons are test particles moving in preset magnetic fields defined over a 3D simulation box, with periodic boundary conditions in all directions. The particles do not interact with each other, as in collisionless plasmas, nor do they induce any fields. Additionally, any radiative energy losses are neglected. An individual electron's motion is, consequently, determined only by the Lorentz force equation given by:
\begin{equation}
\frac{d{\boldsymbol\beta}}{dt} = -\frac{1}{\gamma}\left({\boldsymbol\beta}\times{\boldsymbol\Omega}_B\right),
\label{dvdt}
\end{equation}
where ${\boldsymbol\Omega}_B \equiv e{\bf B}/m_e{c}$. For simplicity, we define our simulation magnetic field as ${\bf B} \equiv {\boldsymbol\Omega}_B$. In this manner, our arbitrary simulation units are always related to a physical magnetic field via the definition of ${\boldsymbol\Omega}_B$. Notice that the purely magnetic Lorentz force conserves particle energy; hence, the velocity vector varies in direction but has a constant magnitude. 

The simulation can be divided into two principle stages (see Ref \citep{keenan12} for a detailed description of the numerical implementation). First, the turbulent magnetic field is created using a predetermined spectral distribution in Fourier space. This field is generated on a cubic lattice that is then interpolated, so as to represent a ``continuous" field. The interpolation implements divergenceless matrix-valued radial basis functions (see Ref. \citep{mcnally11}, for a discussion). This interpolation method starts with a radial function -- in our case, one of the simplest, $\phi({\bf r}) = e^{-\epsilon{r}^2}$ (where $\epsilon$ is a scaling factor, and $r^2 = x^2 + y^2 + z^2$). Then, a set of divergence-free matrix-valued radial basis functions is obtained from the transformation \citep{mcnally11}:
\begin{equation}
\Phi({\bf r}) = (\nabla\nabla^T - \mathbb{I}_{3\times3}{\nabla^2})\phi({\bf r}),
\label{rad_basis}
\end{equation} 
where $\nabla\nabla^T$ is the second-order, $3\times3$-matrix differential operator and $\mathbb{I}_{3\times3}$ is the $3\times3$ identity matrix. 

These interpolants are then applied to the interior of each lattice ``cell'' (the significance of the interpolant's divergence is explored in Appendix \ref{s:appendixb}). The second stage in our model involves the numerical solution of the equation of motion for each particle, i.e.\ Eq. (\ref{dvdt}). From the solution, $\langle\alpha^2\rangle$ and the radiation spectra are obtained. We first turn our attention to the generation of the magnetic field.

As discussed previously (see Ref. \citep{keenan13}), generation of the magnetic field distribution is more convenient in Fourier space. There are two principal reasons for this.

Firstly, it is an easier task to specify a particular spectral distribution in Fourier space directly, rather than attempting to approximate the corresponding field in real space. Secondly, any physically realizable field should satisfy Maxwell's equations, thus its divergence must be zero. This divergenceless condition is more readily met in Fourier space, because Gauss' law, $\nabla\cdot{\bf B}=0$, is an algebraic equation there; ${\bf k}\cdot{\bf B_k}=0$, for each Fourier component. Although our code can handle a wide variety of magnetic spectral distributions, we limit our study to isotropic magnetic turbulence, described in Eq. (\ref{Bk}) -- leaving more sophisticated models for the future. 

After the magnetic field is generated, the next step is the numerical solution of the equation of motion, Eq. (\ref{dvdt}). This was done via a fixed step 4$^\text{th}$-order Runge-Kutta-Nystr\"om method. With all the particle positions, velocities, and accelerations calculated, the radiation spectrum is obtained from Eq. (\ref{LW}).

Next, the total radiation spectrum is obtained by ``summing'' over the spectra of the individual particles. There are two, usually equivalent, methods for doing the summation. First, one can add the spectra coherently by summing over each particle's ${\bf A}_{\bf k}$, and then performing a single integration via Eq. (\ref{LW}). This is a more physical method. In the second method we add the spectra incoherently (i.e., by integrating each particle's ${\bf A}_{\bf k}$ separately, and then summing the results of each integration). As discussed in Ref. \citep{hededal05}, both methods will result in the same spectra, since the wave phases are uncorrelated. However, an incoherent sum will produce spectra that are less noiser, for a given number of simulation particles, than the coherently summed spectra. Hence we use the incoherent approach in our study.

\section{Numerical results}
\label{s:results}

In Section \ref{s:analytic} we made a number of theoretical predictions concerning the transport and radiation properties of plasmas with small-scale turbulent magnetic fields. Additionally, we anticipated that an inter-connection between the transport and radiative properties of non-relativistic/trans-relativistic particles moving through sub-Larmor-scale magnetic turbulence exists, as it does for ultra-relativistic particles \citep{keenan13}. Here we check our predictions, and further explore the radiation spectra.

First of all, we explore how the pitch-angle diffusion coefficient depends on various parameters, cf. Eq. (\ref{Daa_def}), namely the particle's velocity, $\beta$, the magnetic field strength, $\langle B^2 \rangle$, the field correlation scale, $\lambda_B$, and the ``gyro-number'', $\rho$. 

\begin{figure}
\includegraphics[angle = 0, width = 1\columnwidth]{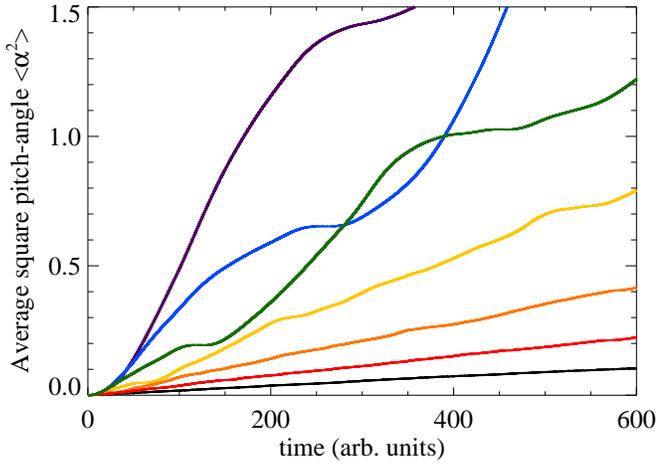}
\caption{(Color online) Average square pitch-angle vs. time (in simulation units). Relevant parameters are $N_p = 2000$, $k_\text{min} = 4\pi/5$, $k_\text{max} = 8\pi$, $\langle B^2 \rangle^{1/2} = 0.01$, and $\mu = 3$. The particle velocities in each case range from $\frac{1}{8}c$ to $\frac{1}{512}c$ (by multiples of two). The curves appear with increasing average slope as $\beta$ decreases. As $\beta$ decreases, eventually $\rho \sim 1$ (at $\beta = \frac{c}{128}$, i.e.\ the fifth most sloped, ``green'' line ), after which the deflection angle becomes large, and pitch-angle diffusion breaks down.}
\label{alpha_break}
\end{figure}

\begin{figure}
\includegraphics[angle = 0, width = 1\columnwidth]{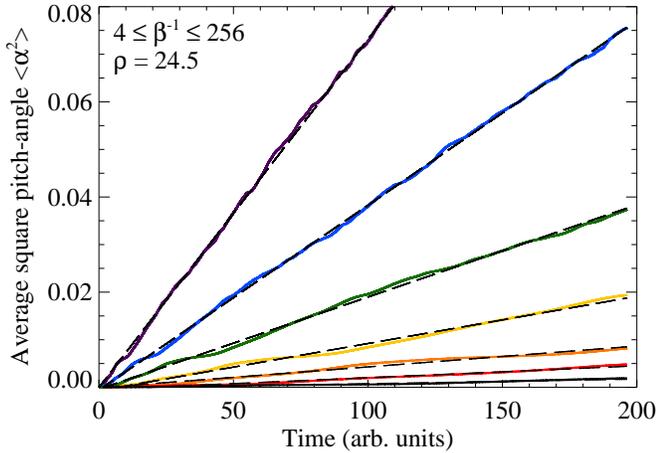}
\caption{(Color online) Average square pitch-angle vs. time (in simulation units). Relevant parameters are $N_p = 2000$, $k_\text{min} = \pi$, $k_\text{max} = 8\pi$, and $\mu = 3$. $\langle B^2 \rangle^{1/2}$ ranges from $5\times10^{-4}$ to $0.032$, by multiples of two. The particle velocities range (in the opposite order) from $\frac{1}{256}c$ to $\frac{1}{4}c$. These two parameters, $\langle B^2 \rangle$ and $\beta$, vary in such a way as to keep $\rho = 24.5$. The curves appear with increasing slope as $\beta$ decreases. Clearly, the linear form of the curves is retained in all seven cases.}
\label{alpha_restore}
\end{figure}

\begin{figure}
\includegraphics[angle = 0, width = 1\columnwidth]{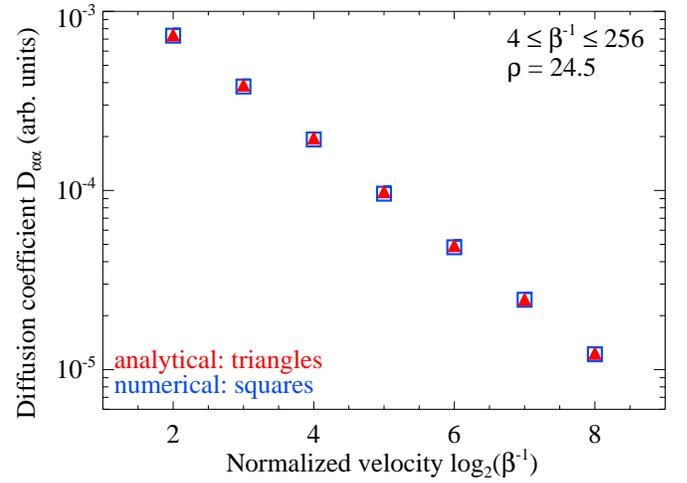}
\caption{(Color online) Pitch-angle diffusion coefficient, $D_{\alpha\alpha}$ vs the logarithm (base 2) of the inverse normalized particle velocity, $log_2(\beta^{-1})$. The (blue) empty ``squares'' indicate the $D_{\alpha\alpha}$ obtained directly from simulation (as the slope of $\langle\alpha^2\rangle$ vs. time), while the (red) filled ``triangles" are the analytical, given by Eq. (\ref{Daa_def}), pitch-angle diffusion coefficients. Simulation parameters are identical to those used in Figure \ref{alpha_restore}.}
\label{diff_v}
\end{figure}

To start, we tested our fundamental assumption that the particle velocity vector only varies slightly over a correlation length, $\lambda_B$. This is the key assumption that underlies our theoretical predictions for pitch-angle diffusion and the radiation spectra. If this assumption were to not hold (i.e.\ if $\alpha_\lambda \ll 1$) then pitch-angle diffusion would break down, and the plot of $\langle \alpha^2 \rangle$ vs time will deviate from linear behavior. In Figure \ref{alpha_break}, $\langle \alpha^2 \rangle$ is plotted as a function of time for seven different cases. In each run, $\langle B^2 \rangle$, $k_\text{min}$, and $N_p$ (number of simulation particles) are fixed to the values of $0.01$, $4\pi/5$ (both in arbitrary simulation units), and $2000$, respectively. The particles are monoenergetic, and are isotropically distributed in their initial velocities. Each case differs in particle velocities; which range from $\frac{1}{512}c$ to $\frac{1}{8}c$. As can be seen, the curves begin as straight lines that increase with slope as $\beta$ decreases. Eventually, the linear behavior breaks down as $\beta$ decreases. A decrease in $\rho$ occurs concurrently, in accordance with Eq. (\ref{scale_para}). As expected, the breakdown in linear behavior, and hence diffusion, occurs when $\rho \sim 1$.
Later, we did the same experiment, only this time we varied $\langle B^2 \rangle$ in such a way as to keep $\rho$ constant ($\rho = 24.5$). In this way, each case is securely in the small-scale regime. In Figure \ref{alpha_restore}, we see that the linear behavior of $\langle \alpha^2 \rangle$ vs time is preserved for all velocities, as anticipated. Consequently, our assumption of a small $\alpha_\lambda$ is valid, as long as $\rho > 1$.
With the existence of pitch-angle diffusion established, we then proceeded to compare the slope of $\langle \alpha^2 \rangle$ vs time (the numerical pitch-angle diffusion coefficient) to Eq. (\ref{Daa_def}). In Figure \ref{diff_v}, the numerically obtained diffusion coefficients from Figure \ref{alpha_restore} are compared to the analytical result of Eq. (\ref{Daa_def}). In each, the theoretical and numerical results differ only by a small factor of ${\cal O}(1)$.

\begin{figure}
\includegraphics[angle = 0, width = 1\columnwidth]{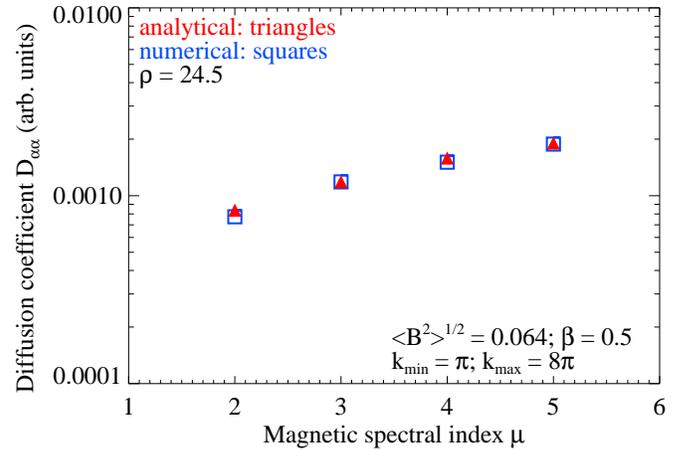}
\caption{(Color online) Pitch-angle diffusion coefficient, $D_{\alpha\alpha}$ vs the magnetic spectral index,  $\mu$. The (blue) empty ``squares'' indicate the $D_{\alpha\alpha}$ obtained directly from simulation, while the (red) filled ``triangles" are the analytical, given by Eq. (\ref{Daa_def}), pitch-angle diffusion coefficients. Relevant parameters are $N_p = 2000$, $k_\text{min} = \pi$, $k_\text{max} = 8\pi$, $\langle B^2 \rangle^{1/2} = 0.064$ , $\beta = 0.5$, and $\rho = 24.5$. The magnetic spectral indexes are $2$, $3$, $4$, and $5$. Notice that the numerical results have nearly the same functional dependence on $\mu$ as the analytical triangles, as given by Eq. \ref{Daa_def}. }
\label{diff_mu}
\end{figure}

Next, we tested the correlation length dependence, i.e.\ whether or not the numerical simulations agree with Eq. (\ref{corr_l}). With $k_\text{min} = \pi$ and $k_\text{max} = 8\pi$, we varied the magnetic spectral index, $\mu$ from $2$ to $5$. This is plotted in Figure \ref{diff_mu}, where the numerical diffusion coefficient closely matches the analytical result.

\begin{figure}
\includegraphics[angle = 0, width = 1\columnwidth]{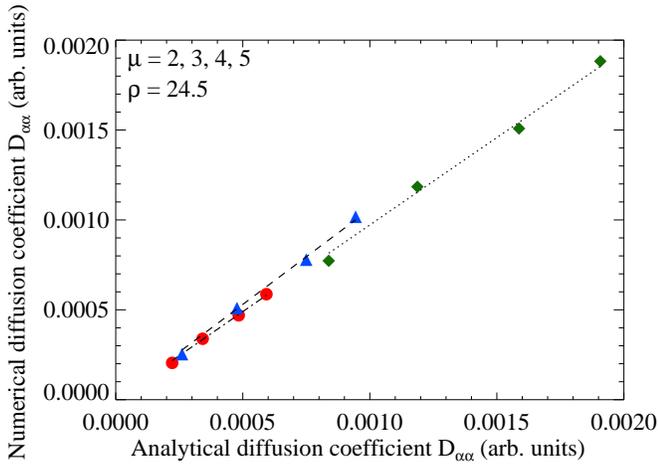}
\caption{(Color online) Numerical pitch-angle diffusion coefficient vs the analytical pitch-angle diffusion coefficient, for three different cases. In each case, the magnetic spectral index has been varied from $2$ to $5$, by intervals of unity. Relevant parameters are $k_\text{min} = \pi/2$ (red) ``circles'' and (blue) ``triangles'', $\pi$ (green) ``diamonds'', $k_\text{max} = 5.12\pi$ (red) ``circles''; $k_\text{max} = 8\pi$ (green) ``diamonds'' and (blue) ``triangles''; $\langle B^2 \rangle^{1/2} = 0.016$ (red) ``circles'', $0.032$ (blue) ``triangles''; $0.064$ (green) ``diamonds''; $\beta = 0.25$ (red) ``circles'', $0.5$ (blue) ``triangles'' and (green) ``diamonds''. In each case, a line of best fit is applied. The slopes are $0.979$ (circles), $0.972$ (diamonds), and $1.06$ (triangles)}
\label{diff_numvsan}
\end{figure}

\begin{figure}
\includegraphics[angle = 0, width = 1\columnwidth]{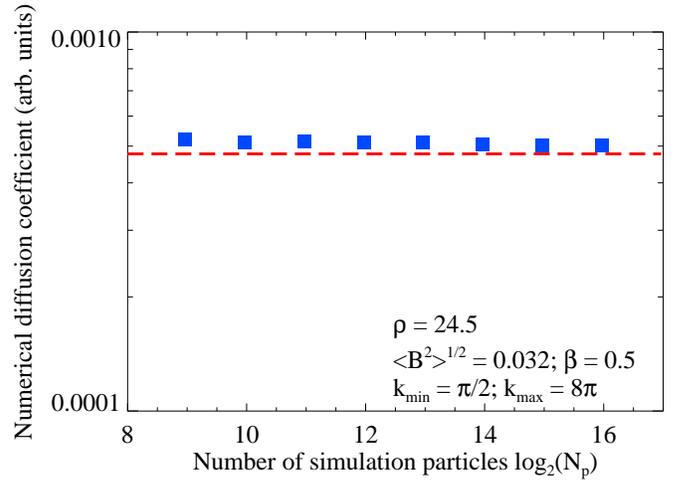}
\caption{(Color online) Pitch-angle diffusion coefficient, $D_{\alpha\alpha}$ vs the total number of simulation particles, $N_p$. The ``blue squares'' indicate the $D_{\alpha\alpha}$ obtained directly from simulation, while the red dotted line is the analytical result, given by Eq. (\ref{Daa_def}). Relevant parameters are $k_\text{min} = \pi/2$, $k_\text{max} = 8\pi$, $\langle B^2 \rangle^{1/2} = 0.032$ , $\beta = 0.5$, and $\rho = 24.5$. There appears to be no strong dependence of the numerical pitch-angle diffusion coefficient upon the total number of simulation particles; nevertheless, there appears to be some convergence to the analytical result.}
\label{diff_num_par}
\end{figure}

\begin{figure}
\includegraphics[angle = 0, width = 1\columnwidth]{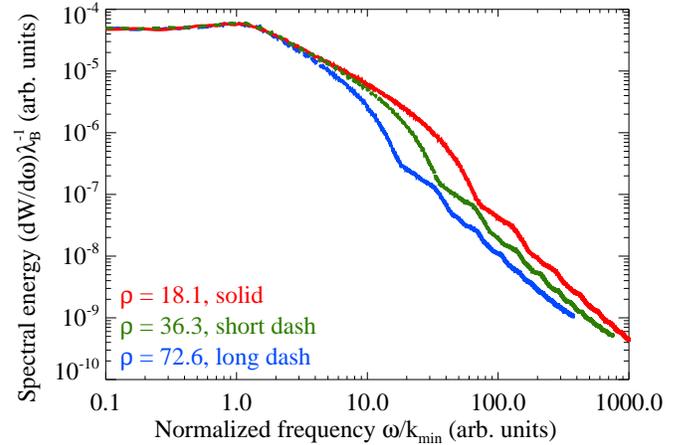}
\caption{(Color online) Radiation spectra given variable $k_\text{min}$, with all other parameters fixed. The number of simulation particles, $N_p$, is $2000$, and $v = 0.5c$ in each case.  In each trial, the particles moved for a total simulation time of $T = T_g$, where $T_g \equiv 2\pi\gamma{m_e}c/e\langle B^2 \rangle^{1/2}$ is the ``gyroperiod''. Here, the axes are in arbitrary, simulation units. We see that the frequency scales as $k_\text{min}$ and $dW/d\omega$ scales as $\lambda_B$.}
\label{kmin_spec}
\end{figure}

\begin{figure}
\includegraphics[angle = 0, width = 1\columnwidth]{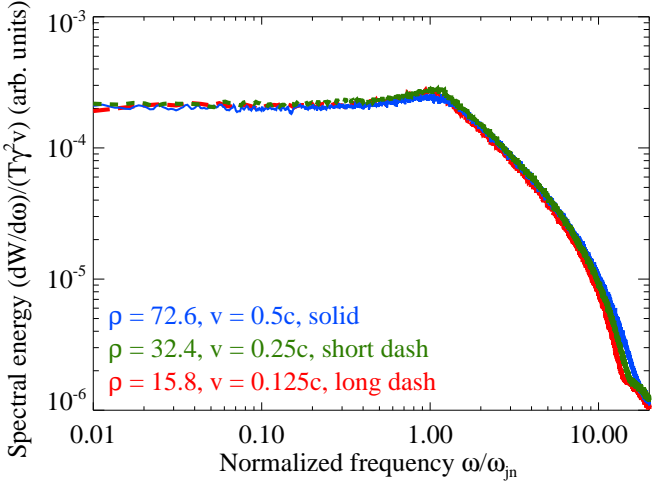}
\caption{(Color online) Radiation spectra given variable $v$. In each trial, $1000$ particles move for a total simulation time of $T = 10T_g$, where $T_g \equiv 2\pi\gamma{m_e}c/e\langle B^2 \rangle^{1/2}$ is the ``gyroperiod''. We see that the overall shape of the spectra is not appreciably altered with decreasing $v$. The spectra are normalized by $T\gamma^2v$, vertically. Given Figure \ref{kmin_spec}, we may conclude that the peak frequency of these spectra is $\omega \sim \gamma^2k_{min}v$ -- cf. Eq. (\ref{omega_jn}).}
\label{v_spec}
\end{figure}

In Figure \ref{diff_numvsan}, the numerical diffusion coefficient is plotted against the analytical coefficient for the same range of $\mu$ values, but now the $k_\text{min}$, $k_\text{max}$, and $\beta$ values differ among the three (with $\rho$ fixed to $24.5$). Included are the results of Figure \ref{diff_mu}. All three cases give a nearly linear relationship between the numerical and analytical coefficients, with slopes approximately equal to unity.
Another concern worth addressing is the dependence of the numerical diffusion coefficient on the total number of simulation particles. In Figure \ref{diff_num_par}, a test case was repeated with an increasing number of simulation particles. The number of particles was increased from $500$ to $64000$, by factors of $2$. There is little variation to be seen in the numerical result, as the number of particles is increased.
Next, we explored the trans-relativistic jitter radiation regime by calculating the radiation spectra, using Eq. (\ref{LW}), with variable simulation parameters. We aimed to test the radiation spectra's dependence upon the key turbulent parameters: $k_\text{min}$, $k_\text{max}$, $\langle B^2 \rangle$, and $\mu$, as well as the particle velocity, $v$. To start, we considered the $k_\text{min}$ dependence. In Figure \ref{kmin_spec}, we have plotted spectra for an initially isotropically distributed, monoenergetic, ensemble of trans-relativistic electrons ($v = 0.5c$) moving through sub-Larmor-scale magnetic turbulence with three different values of $k_\text{min}$. The key parameters are: $\rho = 18.1$, $36.3$, and $72.6$, with $k_\text{min} = \pi/5$, $2\pi/5$, and $4\pi/5$, respectively (see Table \ref{spec_para_table} for a complete listing of simulation parameters used in every figure). The spectra of Figure \ref{kmin_spec}, at least superficially, resemble our theoretical prediction; cf. Eq. (\ref{analy_spec}). We have normalized the $dW/d\omega$ and $\omega$ axes by $\lambda_B$ and $k_\text{min}$, respectively. As expected, the frequency of the spectral peak scales by $k_\text{min}$. 
The precise scaling of the peak frequency is revealed in Figure \ref{v_spec}. In this figure, we have varied the particle velocities, keeping all other parameters fixed. Three velocities appear: $v = 0.125c$, $0.25c$, and $0.5c$. Clearly, the overall shape of the spectra is not strongly dependent upon the particle velocities. We have identified the proper scaling on the horizontal axis. With this result, and Figure \ref{kmin_spec}, we may conclude that the frequency of the peak of the radiation spectrum is $\omega \sim \gamma^2k_{min}v = \omega_{jn}$. This is jitter frequency given in Eq. (\ref{analy_spec}).

\begin{figure}
\includegraphics[angle = 0, width = 1\columnwidth]{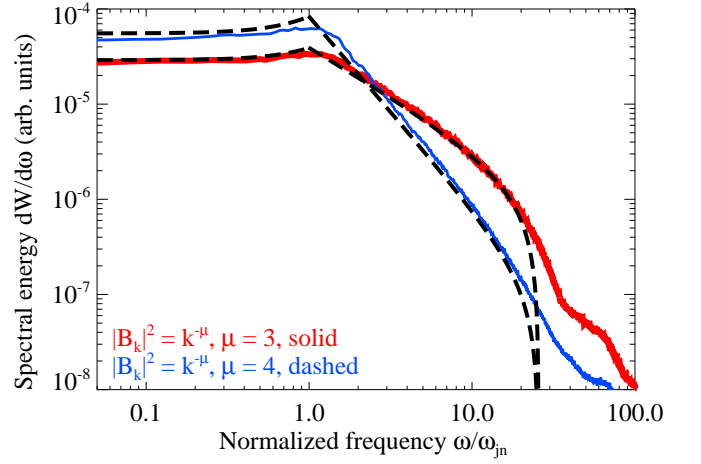}
\caption{(Color online) Radiation spectra given two different values of the magnetic spectral index: $\mu = 5$ (red) ``thick" line and $\mu = 4$ (blue) ``thin" line. Included are the analytical solutions given by Eq. (\ref{analy_spec}). Note that the $\mu = 5$ solution has been multiplied by an overall factor of two for easier visualization. For frequencies near $\omega \sim \gamma^2k_\text{min}v$, the numerical spectra agree decently with the analytical results. However, for frequencies near the break, $\omega \sim \gamma^2k_\text{max}v$, there is considerable deviation between the predicted and numerical spectra -- for both values of the magnetic spectral index. The origin of this discrepancy is explored in Appendix \ref{s:appendixb}. }
\label{mu_spec}
\end{figure}

\begin{figure}
\includegraphics[angle = 0, width = 1\columnwidth]{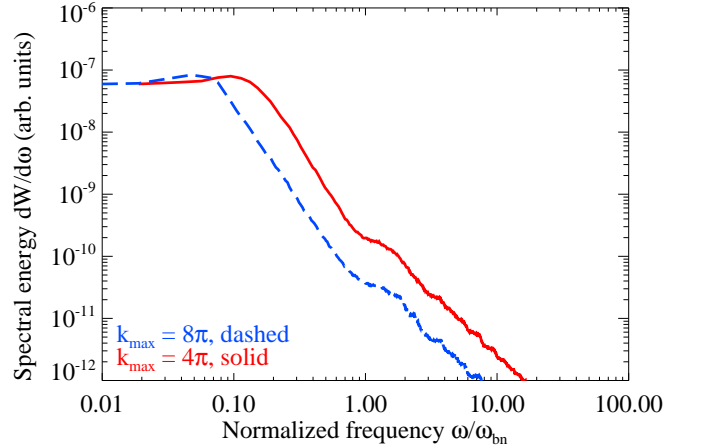}
\caption{(Color online) Radiation spectra with differing $k_\text{max}$. Some other relevant parameters are $v = 0.25c$, $\rho = 6.34$, $N_p = 2000$, and $\mu = 5$ (for a complete listing, see Table \ref{spec_para_table}). The two spectra differ by a factor of 2 in $k_\text{max}$, with $k_\text{min}$ the same between them. Roughly, the spectra transition to the ``tail'' feature near $\omega \sim \gamma^2k_{max}v = \omega_{bn}$.}
\label{kmax_spec}
\end{figure}

\begin{table}
   \centering
   \resizebox{\columnwidth}{!}{%
   \begin{tabular}{*{10}{|c|}r} 
      \toprule
    \midrule
    \hline
     $\#$ & $\rho$ & $\Delta{t}$ & $\beta$ & $\mu$ & $k_\text{min}$ & $k_\text{max}$ & $\sqrt{\langle B^2 \rangle}$ & $N_p$ & $\text{T}_g$ \\
     \hline
     \midrule
$\ref{kmin_spec}$ & $18.1$ & $0.005$ & $0.5$ &  $3$ & $\pi/5$ & $10.24\pi$ &  $0.02$ & $2000$ & $1$    \\ \cline{1-10}
$\ref{kmin_spec}$ & $36.3$ & $0.005$ & $0.5$ &  $3$ & $2\pi/5$ & $10.24\pi$ &  $0.02$ & $2000$ & $1$    \\ \cline{1-10}
$\ref{kmin_spec}$ & $72.6$ & $0.005$ & $0.5$ &  $3$ & $4\pi/5$ & $10.24\pi$ &  $0.02$ & $2000$ & $1$    \\ \cline{1-10}
$\ref{v_spec}$ & $15.8$ & $0.050$ &  $0.125$ & $3$ & $4\pi/5$ & $10.24\pi$ & $0.02$ & $1000$ & $10$ \\  \cline{1-10}
$\ref{v_spec}$ & $32.4$ & $0.050$ &  $0.25$ & $3$ & $4\pi/5$ & $10.24\pi$ & $0.02$ & $1000$ & $10$ \\  \cline{1-10}
$\ref{v_spec}$ & $72.6$ & $0.050$ &  $0.5$ & $3$ & $4\pi/5$ & $10.24\pi$ & $0.02$ & $1000$ & $10$ \\  \cline{1-10}
$\ref{mu_spec}$ & $6.18$ & $0.005$ & $0.125$ & $4$ &  $\pi$ & $8\pi$ & $0.064$ & $8000$ & $1$ \\  \cline{1-10}
$\ref{mu_spec}$ & $6.18$ & $0.005$ & $0.125$ & $5$ &  $\pi$ & $8\pi$ & $0.064$ & $8000$ & $1$ \\  \cline{1-10}
$\ref{kmax_spec}$ & $6.34$ & $0.005$ & $0.25$ &  $5$ & $\pi/2$ & $4\pi$ & $0.064$ & $2000$ & $1$ \\  \cline{1-10}
$\ref{kmax_spec}$ & $6.34$ & $0.005$ & $0.25$ &  $5$ & $\pi/2$ & $8\pi$ & $0.064$ & $2000$ & $1$ \\  \cline{1-10}
$\ref{extreme_mu_spec}$ & $12.4$ & $0.05$ & $0.125$ & $100$ &  $\pi$ & $8\pi$ & $0.032$ & $8000$ & $10$ \\  
\cline{1-10}
$\ref{monopole_spec}$ & $7.9$ & $0.05$ & $0.125$ & $4$ &  $2\pi/5$ & $8\pi$ & $0.02$ & $4000$ & $10$ \\  
\cline{1-10}
$\ref{interp_spec}$ & $6.2$ & $0.005$ & $0.125$ &  $5$ & $\pi$ & $8\pi$ & $0.064$ & $5000$ & $1$ \\  \cline{1-10}
$\ref{op_kdiv10}$ & $14.2$ & $0.00125$ & $0.5$ &  $4$ & $8\pi$ & $400\pi$ & $1.024$ & $1000$ & $10$ \\  \cline{1-10}
$\ref{op_k}$ & $14.2$ & $0.00125$ & $0.5$ &  $4$ & $8\pi$ & $400\pi$ & $1.024$ & $1000$ & $10$ \\  \cline{1-10}
      \bottomrule
   \end{tabular}

}
\caption{Table of parameters used for the radiation spectra figures. Here, and elsewhere, $\Delta{t}$ is the simulation time step, the simulation time is denoted in multiples of the ``gyroperiod'' (i.e.\ $T_g = 2\pi\gamma{m_e}c/e\langle B^2 \rangle^{1/2}$), and $N_p$ is the total number of simulation particles.}
\label{spec_para_table}
\end{table}
Next, we tested the $\mu$ dependence. In Figure \ref{mu_spec}, $\mu = 4, 5$. For each spectrum, $v = 0.125c$, and the total simulation time was $T_g$, where $T_g = e\langle B^2 \rangle^{1/2}/\gamma{m_e}c$ is the gyroperiod. The numerical and analytical spectra show close agreement for frequencies less than the break frequency, $\omega \sim \gamma^2k_\text{max}v$.
In Figure \ref{kmax_spec}, we have plotted two spectra that differ in their $k_\text{max}$ values (all other parameters kept fixed). The $k_\text{max}$ values employed differ by a factor of 2. We see that, roughly, the spectra approach zero near $\omega \sim \gamma^2k_{max}v$. The proceeding power law ``tail'' feature is a numerical artifact that arises from a steep drop to zero power (this fact is more readily apparent in a linear plot -- see Appendix \ref{s:appendixa}).  
Next, we examined the apparent structure in the radiation spectra for $\omega < \omega_{jn}$. This is most clearly seen in Figure \ref{v_spec}, where it appears as a distinctive ``bump''. According to Eq. (\ref{analy_spec}), this bump-like feature has a functional form of $A + D\omega^2$. To assure that this form is correctly identified, we considered a large magnetic spectral index of $\mu = 100$ with $\beta = 0.125c$. Such a large $\mu$ makes the feature more prominent, helping to magnify it. As can be seen, the curve that best fits the bump-like feature at $\omega < \omega_{jn}$ is given by a function of the form $A + D\omega^2$.
\begin{figure}
\includegraphics[angle = 0, width = 1\columnwidth]{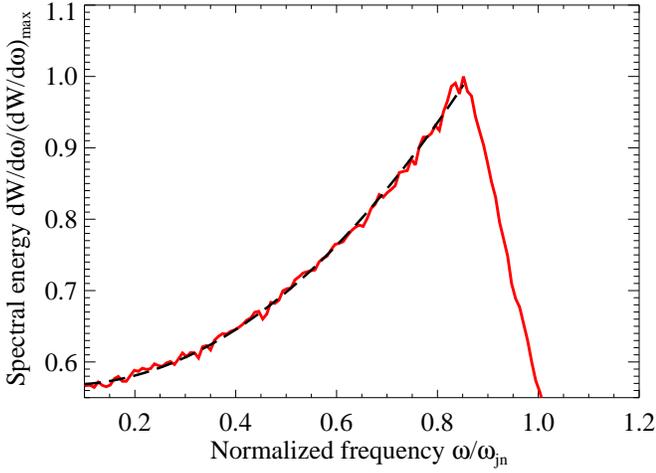}
\caption{(Color online) Radiation spectrum with $\mu = 100$ ($\beta = 0.125c$). Evidently, the spectral feature presented directly prior to $\omega_{jn}$ has a functional form given by $A + D\omega^2$ (dashed line). This is consistent with Eq. (\ref{analy_spec}).}
\label{extreme_mu_spec}
\end{figure}

One may consider the magnetic correlation tensor and its relation to the shape of the radiation spectra. Anisotropic turbulence will alter the shape, but so will a change to the topology of the magnetic field. Motivated by pure curiosity, we consider turbulence that is generated by a distribution of magnetic monopoles. This will result in a magnetic field that is curl-free, but has a divergence given by Gauss's Law for monopoles. This topological change will alter the correlation tensor for isotropic and homogeneous turbulence to \citep{radler11} 
\begin{equation}
B^{i*}_{\bf k}B^{j}_{\bf k}  = \left|{\bf B}_k\right|^2\hat{k}^i\hat{k}^j,
\label{cor_ten_mono}  
\end{equation} 
which is the form required for an irrotational vector field. Substitution of this correlation tensor into Eq. (\ref{dWdw}) will give a slightly different radiation spectrum for the magnetic spectrum in Eq. (\ref{Bk}). 
\begin{figure}
\includegraphics[angle = 0, width = 1\columnwidth]{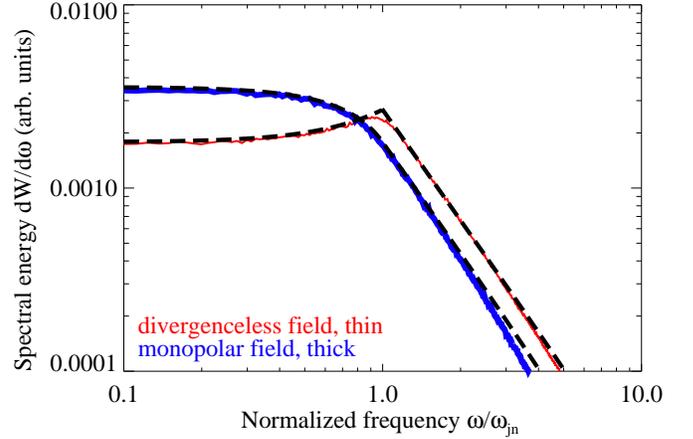}
\caption{(Color online) Radiation spectrum of non-relativistic electrons moving through small-scale magnetic turbulence generated by a distribution of magnetic monopoles (``thick'', blue), superimposed with the radiation spectrum given a magnetic spectrum (``thin'', red) produced by standard means (i.e.\ Ampere's Law). For each run, $\mu = 4$ and $\beta = 0.125c$. Each curve is accompanied by its corresponding analytical solutions (``dashed'', black). The spectral shape for frequencies less than $\omega_{jn}$ is $A + D\omega^2$ and $A - D\omega^2$ for the ``divergenceless'' field and ``monopolar'' field, respectively.}
\label{monopole_spec}
\end{figure}
The principal change will be to the quadratic prefactor $A + D\omega^2$. The ``monopolar'' field will result in a sign change to $D$. In Figure \ref{monopole_spec}, this difference is clearly indicated. Notice the apparent lack of the quadratic peak feature at $\omega_\text{jn}$. 

\begin{figure}
\includegraphics[angle = 0, width = 1\columnwidth]{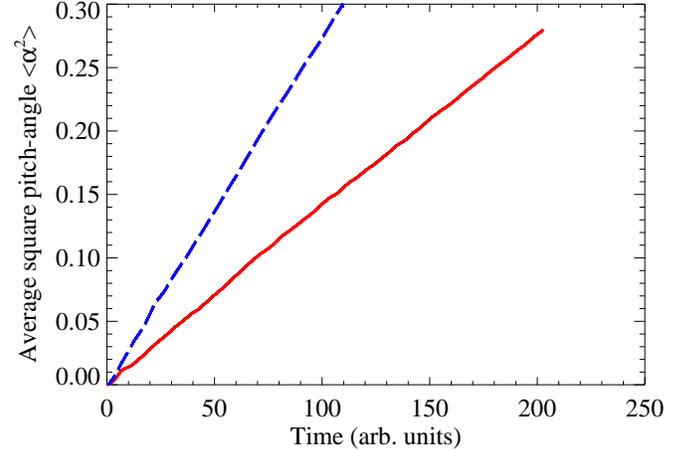}
\caption{(Color online) Average square pitch-angle growth as a function of time for non-relativistic electrons moving through small-scale magnetic turbulence generated by a distribution of magnetic monopoles ``dashed'' (blue), superimposed with the otherwise equivalent curve ``solid'' (red) produced by standard means (i.e.\ Ampere's Law). For each run, $\mu = 6$, $N_p = 15420$, $\langle B^2 \rangle^{1/2} = 0.032$, $k_\text{min} = \pi$, $k_\text{max} = 8\pi$, and $\beta = 0.125c$. Note that the slope of the ``monopolar'' curve is very nearly twice the slope of the standard curve -- in accordance with Eq. (\ref{diff_mono_cor}).}
\label{alpha_mono}
\end{figure}

The altered correlation tensor will affect the particle diffusion coefficient as well. In fact, as can be seen in Figure (\ref{alpha_mono}), the pitch-angle diffusion coefficient of particles moving in the monopolar field is twice as large as the divergenceless field equivalent. This follows from the fact that
\begin{equation}
\lambda_B^\text{monopole} = 2\lambda_B^\text{div. free},
\label{diff_mono_cor}  
\end{equation} 
which results from substitution of Eq. (\ref{cor_ten_mono}) into Eq. (\ref{corr_l_def}).
It is a noteworthy observation that the preceding results are identical, up to overall multiplicative factors, to the radiation spectra and pitch-angle diffusion coefficient for the more physically plausible situation of a trans-relativistic monopole moving through ``small-scale'' electrostatic turbulence, such as Langmuir turbulence. 

\section{Conclusions}
\label{s:concl}

In this paper we explored non-relativistic and trans-relativistic particle transport (diffusion) and radiation production in small-scale electromagnetic turbulence. Principally, we demonstrated that in the regime of small deflections, i.e.\ when the particle's deflection angle over a correlation length is small $\alpha_\lambda \ll 1$, the pitch-angle diffusion coefficient and the simultaneously produced radiation spectrum are wholly determined by the particle velocity and the statistical/spectral properties of the magnetic turbulence; which is a result most transparently offered by Eqs. (\ref{corr_l_div}) and (\ref{nonrel_analy}). Additonally, we showed that the condition of a small deflection angle is satisfied if $\rho > 1$, i.e.\ if the magnetic turbulence is small-scale. 

These results generalize the ultra-relativistic regime first discussed in Ref. \citep{keenan13}. In fact, the pitch-angle diffusion coefficient remains unchanged, in both the non-relativistic and relativistic regimes. Significantly, just as small-angle jitter radiation strongly differs from synchrotron radiation, so too does the analogous non-relativistic jitter radiation distinguish itself from cyclotron radiation. Given the isotropic 3D power law magnetic spectral distribution from Eq. (\ref{Bk}), the resulting trans-and non-relativistic radiation spectrum is a piece-wise function of a quadratic equation in frequency, $\omega$ up to the characteristic (jitter) frequency, $\omega_{jn} = \gamma^2k_\text{min}v$, after which it is the sum of a power law and a quadratic term up to the break frequency, $\omega_{bn} = \gamma^2k_\text{max}v$, where it then quickly approaches zero -- see Eq. (\ref{analy_spec}).
We have, further, confirmed our theoretical results via first-principle numerical simulations. 

Lastly, we have considered the change in the radiative and transport properties of trans-relativistic particles moving through magnetic turbulence due to a topological change in the field. Namely, we supposed the generation of sub-Larmor-scale magnetic turbulence from a distribution of magnetic monopoles. We showed that the radiation spectra and pitch-angle diffusion coefficient are modified; i.e.\ the pitch-angle diffusion coefficient doubles in magnitude, \emph{\`{a} la} Eq. (\ref{diff_mono_cor}), and the shape of the radiation spectrum is dramatically altered for frequencies less than the jitter frequency, $\omega_{jn}$. These results, furthermore, generalize to the case of a magnetic monopole moving through ``small-scale" electrostatic turbulence.

Finally, the applicability of our model will depend heavily upon the plasma environment. The turbulence dissipation time-scale, growth rate, time-evolution, and spatial-scale are important considerations. We have highlighted the Weibel-like turbulence, in particular, because of its favorable properties. As stated previously, the Weibel instability can produce strong, small-scale, magnetic fields in an non-magnetized plasma. Furthermore, the instability is aperiodic (i.e.\ real frequency $\Omega_{r} \sim 0$), and thus allows for the static field treatment. More precisely, the growth rate $\gamma \gg \Omega_{r}$. Typically, the growth rate is governed by a characteristic plasma frequency. Lastly, the magnetic fluctuations are long-lived in the case of the Weibel-filamentation instability, dying out only when the driving free energy (provided by the kinetic energy of streaming particle filaments) of the system is converted by particle isotropization (i.e.\ the depletion of the anisotropy in the streaming particle distribution function). In short, the generated fields are approximately stationary on a time-scale which exceeds the growth/stabilization rate times \citep{treumann12}. Consequently, there appears to be adequate time for radiation production in the jitter regime, given by our prescription, in these ``quasi-static'' Weibel magnetic fields.

Via subsequent non-linear evolution, the electron-generated Weibel magnetic fields may grow to larger spatial-scales -- including the ion skin-depth. Additionally, the Weibel fields may ``seed'' the growth of further MHD turbulence via a process of inverse-cascade -- once more, residing at larger spatial-scales. Thus, in the non-relativistic regime, the jitter radiation spectrum may be effectively screened out when the turbulent magnetic fields predominantly exist at scales much larger than the electron skin-depth. Consequently, non-relativistic jitter radiation, as a diagnostic of Weibel turbulence, may have a limited applicability. However, kinetic instabilities in magnetized plasma can produce turbulent magnetic spectra at the appropriate length scales as well. One such scenario may be provided by a turbulent magnetic field generated in a cold, magnetized, background plasma. We then imagine the existence of a ``hot'' population of sub-Larmor-scale electrons that will serve as our test particles. To this end, anisotropic whistler turbulence may provide a promising candidate. In fact, the (low beta) collisionless Whistler spectrum (perpendicular to the mean magnetic field) may be rather broadband -- a (stationary) piece-wise set of power-laws extending to scales much smaller than the electron skin-depth \citep{saito10}. Naturally, our model requires modification to suit a magnetized plasma -- the case to be considered elsewhere.

To conclude, the obtained results, coupled with our previous work, reveal strong inter-relation of transport and radiative properties of plasmas turbulent at sub-Larmor scales -- whether they be relativistic or non-relativistic. We have demonstrated how spectral information can be a powerful tool to diagnose magnetic micro-turbulence in laboratory and astrophysical plasmas.

\begin{acknowledgments}
This work has been supported by the DOE grant DE-FG02-07ER54940 and the NSF grant AST-1209665.
\end{acknowledgments}

\appendix

\section{The spectral tail} 
\label{s:appendixa}

As can be seen in Figure \ref{kmin_spec} and Figure \ref{kmax_spec}, there is additional structure to the radiation spectra beyond the break frequency, $\sim \gamma^2k_{\text{max}}v$.  This feature is, in fact, a numerical artifact that is magnified by the use of a log-log plot. Here we have plotted Figure \ref{kmin_spec} on a linear scale, and have normalized the frequency axis by the spectral break frequency $\omega_{bn} = \gamma^2k_{\text{max}}v$. 

\FloatBarrier
\begin{figure}
\includegraphics[angle = 0, width = 1\columnwidth]{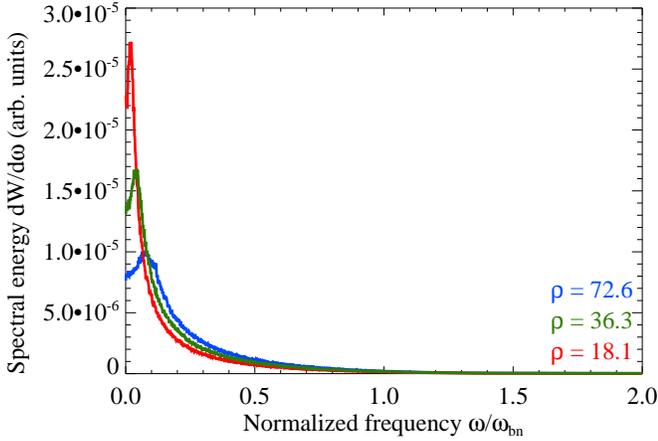}
\caption{(Color online) Radiation spectra of Figure \ref{kmin_spec}, with linear abscissa. We see that the power spectrum quickly approaches zero around the ``break'' frequency, $\gamma^2k_{max}v$ -- in accord with Eq. (\ref{analy_spec}). This numerical approach to zero, since it is not instantaneous, appears readily in a log-log plot which magnifies features on an orders of magnitude scale.}
\label{tail}
\end{figure}
\FloatBarrier

\section{Interpolation of the magnetic field}
\label{s:appendixb}

\begin{figure}[H]
\includegraphics[angle = 0, width = 1\columnwidth]{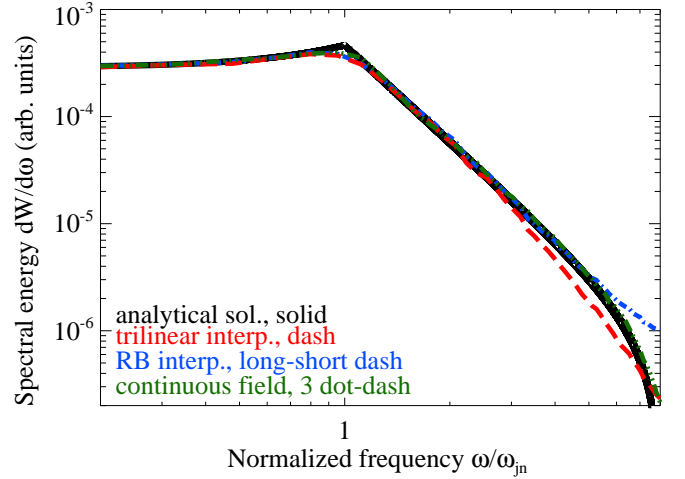}
\caption{Radiation spectra given two different interpolations of the magnetic field and a ``continuous'' field. Relevant parameters are $v = 0.125c$, $\rho = 24.7$, $N_p = 2000$ (for a complete listing, see Table \ref{spec_para_table}). The number of wave modes employed to produce the ``continuous'' magnetic field was $N_m = 10000$. For small frequencies, there is little deviation between the spectra. It is only near the ``break'' frequency (i.e.\ $\omega_{bn} = \gamma^2k_{max}v$) that the three differ considerably. Both of the interpolation derived spectra largely deviate from the analytical solution at the high frequency end; however, the ``continuous'' field derived spectrum differs noticeably only at the outermost frequencies. Whether or not this deviation is solely to blame on the quality of the interpolant or the discrete nature of lattice derived field, has let to be determined. At any rate, both interpolants fail to preserve the slope of the spectra up to $\omega_{bn}$, and there is considerable difference between the divergence-free and trilinear cases. }
\label{interp_spec}
\end{figure}

One might consider the importance of using a divergenceless set of interpolants for the magnetic field. In Figure \ref{interp_spec}, we show a spectrum obtained via the divergenceless radial-basis interpolants of Eq. (\ref{rad_basis}) with a spectrum obtained using a simple, non-divergenceless, trilinear interpolation of the magnetic field. For small frequencies, there is little disagreement between the two spectra. However, as the curves approach the break frequency $\omega_{bn} = \gamma^2k_{max}v$, considerable deviation between the trilinear and radial basis interpolants occurs. In our previous work on the relativistic small-angle jitter regime (see Ref. \citep{keenan13}), little deviation in these spectra was observed in our test runs. One possible explanation is that, since the particle velocities were $\sim \text{c}$, the total distance traveled by a particle in one time step was $\Delta{x} \sim c\Delta{t}$. The spacing between lattice points is, typically, within an order of $c\Delta{t}$. In this case, the interpolant should not play an important role in determining the particle trajectories. If, however, $v$ is much less than $c$, then the difference may be significant. In Figure \ref{interp_spec}, $v = 0.125c$, thus $\Delta{x} \sim 0.125c\Delta{t}$ (an order of magnitude smaller). In this case, frequencies in the radiation spectrum at scales comparable to the grid resolution (i.e.\ large $\text{k}$'s) will suffer the most from this deviation.
Another question worth addressing is the influence of the discrete implementation of the magnetic field on the spectral shape. Recall that the random magnetic field is initially generated on a lattice in $k$-space, after which it is subsequently transformed by FFT to real space. The interpolation is then applied on the lattice of points. Due to memory limitations, the lattice dimensions are limited to $\sim 500^3$; this can be a very severe limitation on the spatial resolution of the magnetic field.

An alternative generation of the magnetic field -- which is grid-less and, therefore, not requiring interpolation -- employs a large sum of sinusoidal wave modes which are evaluated at each time step (as needed). Thus, the magnetic field is effectively ``continuous'' in this representation. Each wave mode is constructed with a random phase and random polarization vector (which is constrained to the plane perpendicular to ${\bf k}$; thus satisfying Gauss's law). The polarization vector may be generated by a variety of methods, but we have chosen the implementation described by Ref. \citep{tautz}. This representation of the polarization vectors is designed specifically to simultaneously satisfy the required properties of isotropic, homogeneous, and divergence-free magnetic turbulence. Additionally, the wave numbers, ranging from $k_\text{min}$ to $k_\text{max}$, are logarithmically spaced. 

In Figure \ref{interp_spec}, we also included a radiation spectrum obtained by electrons moving in the ``continuous'' magnetic field (with, otherwise, identical properties). Evidently, the ``continuous'' field derived spectrum closely matches the analytical solution, Eq. (\ref{analy_spec}) -- preserving the high-frequency end better than the interpolation derived spectra.

\section{Comment on pitch-angle diffusion in the ultra-relativistic regime}
\label{s:appendixc}

We wish to address an error in our previous paper on relativistic pitch-angle diffusion in sub-Larmor-scale magnetic turbulence, Ref. \citep{keenan13}. The paper contains a table for a plot (Figure $7$) of the diffusion coefficient vs the corresponding radiation spectral peak, for relativistic particles moving through a small-scale magnetic field. The magnetic field has identical properties to those employed in this paper. The table contains some errors, which we address here by providing a corrected table (see Table \ref{old_correc}).
\newline
\begin{table}[H]
   \centering
    \resizebox{\columnwidth}{!}{%
   \begin{tabular}{*{9}{|c|}r} 
      \toprule
    \midrule
    \hline
     $\#$ & $\delta_j$ & $\Delta{t}$ & $\gamma$ & $\mu$ & $k_\text{min}$ & $k_\text{max}$ & $\langle B^2 \rangle^{1/2}$ & $N_p$ \\
     \hline
     \midrule

$1$ & $0.63$ & $0.0100$ & $8$ & $3$ &  $1.0$ & $16.1$ & $0.100$ & $2000$ \\ \cline{1-9} 
$2$ & $0.47$ & $0.0100$ & $7$ & $3$ &  $0.6$ & $16.1$ & $0.047$ & $500$  \\ \cline{1-9}
$3$ & $0.12$ & $0.0025$ & $5$ & $3$ &  $0.6$ & $32.2$ & $0.047$ &  $4000$   \\ \cline{1-9} 
$4$ & $0.47$ & $0.0100$ & $3$ & $3$ &  $0.6$ & $16.1$ &  $0.047$ & $500$ \\  \cline{1-9} 
$5$ & $0.94$ & $0.0100$ & $5$ & $3$ &  $0.3$ & $16.1$ & $0.047$ & $500$  \\ \cline{1-9}
      \bottomrule
   \end{tabular}
}
\caption{Corrected table of parameters used in Figure $7$ of ref. \citep{keenan13}, and Figure (\ref{diff_peak_correc}). The correction is as follows: $\#2 \rightarrow \#1$, $\#3 \rightarrow \#2$, and $\#1 \rightarrow \#3$; in what was previously $\#1$, $k_\text{min}$ has been changed from $1.3$ to $0.6$ and $\langle B^2 \rangle^{1/2}$ has been changed from $0.024$ to $0.047$.}
\label{old_correc}
\end{table}

Additionally, we have opted to reproduce Figure $7$ from Ref. \citep{keenan13}, to recalculate the analytical pitch-angle diffusion coefficient. In our previous paper, we used Eq. (\ref{Daa}), as we have here, but with cruder approximations for $\lambda_B$ and $\langle \beta_\perp^2 \rangle^{1/2}$ -- namely, $\langle \beta_\perp^2 \rangle^{1/2} \approx 1$ and $\lambda_B \approx k_\text{min}^{-1}$. 

\begin{figure}[H]
\includegraphics[angle = 0, width = 1\columnwidth]{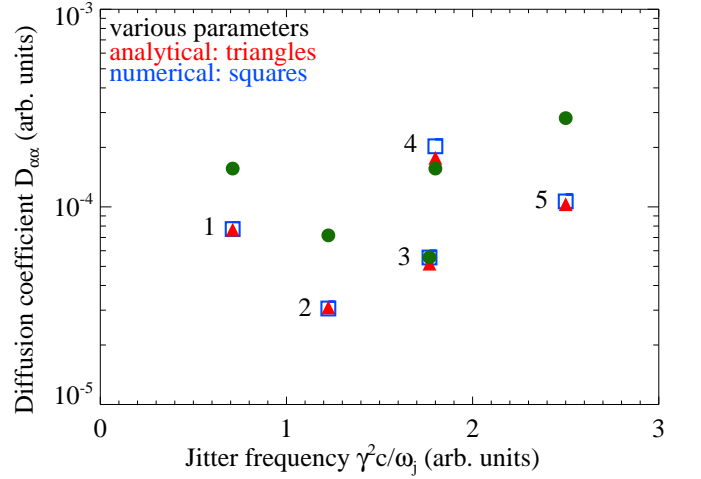}
\caption{(Color online) Modified figure of $D_{\alpha\alpha}$ vs the frequency of $\omega_j = \gamma^2k_\text{min}c$, from Ref. \citep{keenan13}. Once more, the (blue) empty ``squares" indicate the $D_{\alpha\alpha}$ obtained directly from simulation while the (red) filled ``triangles" are the analytical $D_{\alpha\alpha}$, given by Eq. (\ref{Daa_def}). The analytical solution from Ref. \citep{keenan13} appears as (green) filled ``circles''. Notice that the redefined analytical $D_{\alpha\alpha}$'s (red) empty ``triangles'' eliminate the wider variance seen in the cruder approximation (green) filled ``circles".}
\label{diff_peak_correc}
\end{figure}

Now, the refined definition for $D_{\alpha\alpha}$, Eq. (\ref{Daa_def}), eliminates the wider variance between the theoretical and numerical results (see Figure \ref{diff_peak_correc}). There continues to exist a small difference between the analytical and numerical pitch-angle diffusion coefficients, but this variation is relatively small in each case; despite the variability in the simulation parameters employed.

\section{The effect of plasma dispersion on the radiation spectra}
\label{s:appendixc}

As mentioned in Section \ref{s:analytic}, inclusion of plasma dispersion changes the non-relativistic radiation spectrum to
\begin{equation}
\frac{d^2W}{d\omega\, d\eta} = 
\frac{e^2}{4\pi^2 c^3}  \sqrt{\epsilon(\omega)}\left|{\bf w}_{\omega}\right|^2{sin^2\Theta},
\label{LW_nonrel_disp}  
\end{equation} 
where $\epsilon(\omega) = 1 - \omega_\text{pe}^2/\omega^2$, is the plasma scalar permittivity. Since this amounts to a multiplicative factor, the jitter spectrum Eq. (\ref{analy_spec}) will be modified simply by an extra frequency-dependent coefficient. The effect will, however, further complicate the relativistic regime. Fortunately, a Lorentz transformation can be applied, once more, to obtain the relativistic spectrum. 

\indent
Consider a relativistic electron moving with velocity $\beta$ in the (unprimed) plasma rest frame. In this frame, the plasma frequency is $\omega_\text{pe}$; additionally, the index of refraction is $n \equiv \sqrt{\epsilon}$. Conversely, the electron rest frame will be the site of a plasma in motion, with velocity $-\beta$. In this frame, $\omega_\text{pe}' = \omega_\text{pe}/\sqrt{\gamma}$. In a plasma medium, the radiation spectra are connected by the generalized relation 
\begin{equation}
\frac{1}{n\omega^2}\frac{d^2W}{d\omega{d\eta}} = \frac{1}{n'\omega'^2}\frac{d^2W'}{d\omega'{d\eta'}},
\label{spec_invar_disp}  
\end{equation} 
where $n'$ is the index of refraction in the electron rest frame. Via Lorentz transformation, $n'$ is \citep{elitzur}
\begin{equation}
{n'}^2 - 1 = (\omega/\omega')^2(n^2 - 1),
\label{ind_transform}  
\end{equation} 
from which one may obtain the generalization of the relativistic Doppler effect 
\begin{equation}
\omega' = \gamma\omega(1 - {\bf N}\cdot{\boldsymbol\beta}),
\label{doppler_disp}  
\end{equation} 
where ${\bf N} \equiv n\hat{\bf n}$. Using the reverse transformation, i.e.\ $\emph{prime} \leftrightarrow \emph{unprimed}$ and $\beta \rightarrow -\beta$, the angle cosines are related by
\begin{equation}
cos\theta' = \frac{n{cos\theta}-\beta}{n'(1-n\beta{cos\theta})}.
\label{cos_disp}  
\end{equation} 
Using these results, along with Eqs. (\ref{angle_transform}) and (\ref{jitter_vel_free}), the dispersion corrected relativistic jitter spectrum becomes
\begin{equation}
\frac{dW}{d\omega} = \frac{3n}{8\gamma^2} \int_{-1}^1 \! \mathrm{d} {x} \left[\frac{1}{(1 - n\beta{x})^2} + \frac{(nx-\beta)^2}{{n'}^2(1-n\beta{x})^4}\right]I(\omega_0),
\label{jitter_vel_free_disp}  
\end{equation} 
with $\omega_0 \equiv \omega(1-n\beta{x})$ and
\begin{equation}
 n' = \frac{\sqrt{n^2 - 1 + \gamma^2(1 - n\beta{x})^2}}{\gamma(1-n\beta{x})}.
\label{n_prime_def}  
\end{equation} 
\indent
Next, the numerical spectrum is obtained from the generalizations of Eqs. (\ref{LW}) and (\ref{A_k})
\begin{equation}
\frac{d^2W}{d\omega\, d\eta} = 
\sqrt{\epsilon(\omega)}\frac{e^2}{4\pi^2 c}  \left|\int_{-\infty}^\infty \! {\bf A}_{\bf k}(t)e^{i\omega{t}}\, \mathrm{d} t
\right|^2,
\label{LW_disp}  
\end{equation} 
where
\begin{equation}
{\bf A}_{\bf k}(t) \equiv \frac{\hat{\bf n}\times[(\hat{\bf n} - {\boldsymbol\beta}) \times \dot{\boldsymbol\beta} ]}{(1 - \sqrt{\epsilon(\omega)}\hat{\bf n}\cdot{\boldsymbol\beta})^2}e^{-i\sqrt{\epsilon(\omega)}{\bf k}\cdot {\bf r}(t)}.
\label{A_k_disp}  
\end{equation} 

\begin{figure}[H]
\includegraphics[angle = 0, width = 1\columnwidth]{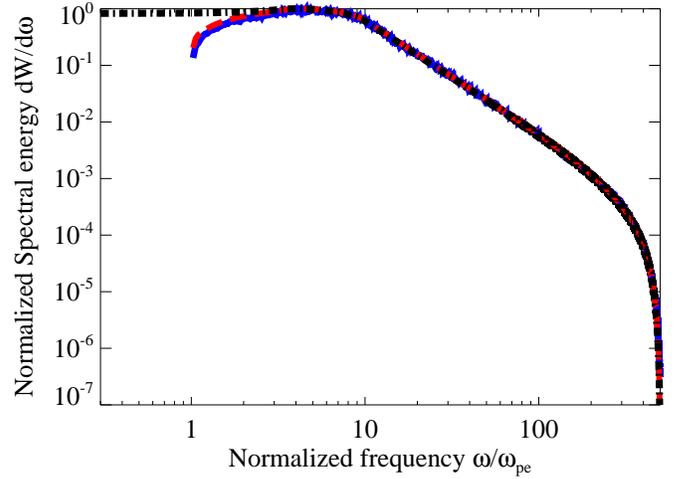}
\caption{(Color online) Numerical radiation spectrum given a $\beta = 0.5$ electron moving through sub-Larmor-scale magnetic turbulence in a dispersive plasma (``thick'', blue), superimposed with the analytical spectrum from Eq. (\ref{jitter_vel_free_disp}) (``dashed'', red) and the ``dispersion-free'' spectrum (``long-short dash'', black). For these runs, $\mu = 4$, $\rho = 14.2$, and $\omega_\text{pe} = k_\text{min}c/10$ (see Table \ref{spec_para_table} for a complete listing of simulation parameters). All spectra are normalized to their respective maximum values. As can be readily seen, the high-frequency end remains largely unchanged by the inclusion of plasma dispersion.}
\label{op_kdiv10}
\end{figure}

In Figure \ref{op_kdiv10}, we consider a $\beta = 0.5$ electron moving through a plasma medium with a plasma frequency $\omega_\text{pe} = k_\text{min}c/10$. The plot includes the equivalent dispersion-free jitter spectrum, along with the analytical spectrum, from Eq. (\ref{jitter_vel_free_disp}), and a spectrum obtained numerically. The numerical spectrum was produced given magnetic turbulence prescribed by the model described in Appendix \ref{s:appendixb}. Since the wave number becomes imaginary when $\omega < \omega_\text{pe}$, we have set a cut-off for frequencies below the plasma frequency. From the plot, we see that the spectrum differs largely from the dispersion-free equivalent for frequencies near $\omega_\text{pe}$. However, as anticipated, the high-frequency end is largely unaffected. 
\begin{figure}[H]
\includegraphics[angle = 0, width = 1\columnwidth]{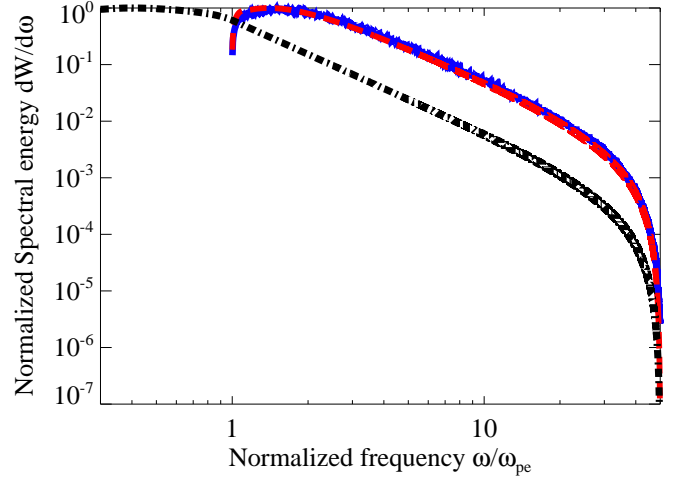}
\caption{(Color online) Radiation spectra, identical to Figure \ref{op_kdiv10},  with the exception that $\omega_\text{pe} = k_\text{min}c$. With $\omega_\text{pe} \sim \omega_{jn}$, the dispersion plays a more prominent role. Nonetheless, the overall shape of the spectrum is unaffected.}
\label{op_k}
\end{figure}

\indent
However, as can be seen in Figure \ref{op_k}, the spectrum is altered in a more dramatic way when $\omega_\text{pe} = k_\text{min}c$. The low-frequency end remains distinctly concave, but now the high-frequency end is shifted to the right. The overall shape of the spectrum, nevertheless, remains the same.

\begin{figure}[H]
\includegraphics[angle = 0, width = 1\columnwidth]{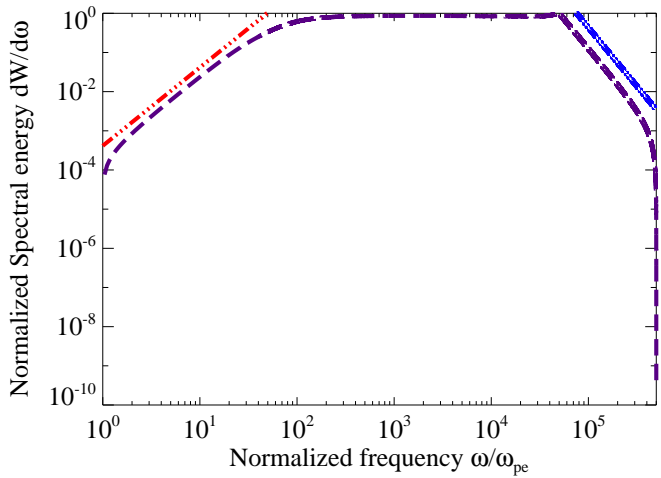}
\caption{(Color online) Dispersion adjusted analytical radiation spectrum for a $\gamma = 50$ electron. Relevant parameters are $\rho = 153.4$ and $\mu = 5$. Two power laws appear. The $\omega^2$ (``long-three-dash'', red) power law, which extends up to $\omega \sim \gamma\omega_{pe}$, is a consequence of the Razin effect. Additionally, we have included $\omega^{-\mu + 2}$ (``long-two-dash'' blue) on the right. As expected from Eq. (\ref{Pomega}), the high-frequency end is a power law, with a very steep drop beyond $\omega_{bn} \approx \omega_b$.} 
\label{disp_gam_50}
\end{figure}

\indent
As a final test of Eq. (\ref{jitter_vel_free_disp}), we consider an extreme relativistic case, $\gamma = 50$. The ultra-relativistic jitter spectrum, with plasma dispersion included, contains an additional $\omega^2$ asymptote at low-frequencies (a hint of this was seen in the previous, trans-relativistic, plots). In Figure \ref{disp_gam_50}, we see the emergence of this low-frequency asymptote. Additionally, we see that the jitter spectrum falls off dramatically for frequencies beyond $\omega_{jn} = \gamma^2k_\text{max}v \approx \gamma^2k_\text{max}c$ -- hence, the correspondence to the ultra-relativistic hard cut-off at $\omega_b$, from Eq. (\ref{Pomega}), is made apparent.

\pagebreak

\end{document}